\documentclass[11pt,a4paper]{amsart}
\usepackage[margin=2.2cm,top=1.8cm,bottom=1.8cm]{geometry}
\usepackage{amsmath,amssymb,mathtools}
\usepackage{tikz}
\usepackage{enumitem}
\usepackage{bm}
\numberwithin{equation}{section}
\setlist[enumerate]{leftmargin=2.5em,itemsep=5pt,topsep=6pt,parsep=3pt}
\title{Matrix Star-Triangle Relations}
\author{Rinat Kashaev}
\address{Section de Mathématiques, Université de Genève,
rue du Conseil-G\'en\'eral 7-9, 1205 Gen\`eve, Suisse\\}
\email{Rinat.Kashaev@unige.ch}
\date{Started: November 5, 2026; updated: \today}
\newcommand{\id}{\operatorname{id}}
\newcommand{\ev}{\operatorname{ev}}
\newcommand{\coev}{\operatorname{coev}}
\newcommand{\dw}{\dot{w}}
\newcommand{\ri}{\mathrm{i}}
\newcommand{\Z}{\mathbb{Z}}
\newcommand{\zm}[1]{\Z/#1\Z}
\newcommand{\rc}{\mathcal{C}}

\newcommand{\calK}{\mathcal K}
\newcommand{\ram}{\Psi}

\newcommand{\diagW}[2]{%
  \mathord{\vcenter{\hbox{%
    \begin{tikzpicture}[x=1bp,y=-1bp,line cap=butt,line join=miter]
      \path[draw=black,line width=0.797010bp] (27.34607,30.51925) -- (2.90341,6.07659);
      \path[draw=black,line width=0.563565bp,fill=black] (1.50266,4.67574) -- (2.67813,7.29134) -- (2.76260,5.93569) -- (4.11826,5.85122) -- cycle;
      \path[draw=white,line width=3.586600bp] (0.70001,30.51925) -- (25.16129,6.05797);
      \path[draw=white,line width=0.608841bp,fill=white] (26.47906,4.74020) -- (24.05182,5.82897) -- (25.31351,5.90575) -- (25.39029,7.16744) -- cycle;
      \path[draw=black,line width=0.797010bp] (0.70001,30.51925) -- (25.14267,6.07659);
      \path[draw=black,line width=0.563565bp,fill=black] (26.54352,4.67574) -- (23.92792,5.85122) -- (25.28357,5.93569) -- (25.36804,7.29134) -- cycle;
      \node[anchor=base west,inner sep=0pt,outer sep=0pt] at (11.14026,9.15282) {\fontsize{10.95pt}{13.6pt}\selectfont $#1$};
      \node[anchor=base west,inner sep=0pt,outer sep=0pt] at (11.68225,31.37578) {\fontsize{10.95pt}{13.6pt}\selectfont $#2$};
      \pgfresetboundingbox
      \path[use as bounding box] (0,0) rectangle (28.04608,34.10305);
    \end{tikzpicture}%
  }}}%
}

\newcommand{\diagWminus}[2]{%
  \mathord{\vcenter{\hbox{%
    \begin{tikzpicture}[x=1bp,y=-1bp,line cap=butt,line join=miter]
      \path[draw=black,line width=0.797010bp] (0.70000,30.51919) -- (25.14265,6.07653);
      \path[draw=black,line width=0.563565bp,fill=black] (26.54352,4.67565) -- (23.92792,5.85113) -- (25.28357,5.93560) -- (25.36804,7.29125) -- cycle;
      \path[draw=white,line width=3.586600bp] (27.34605,30.51919) -- (2.88477,6.05791);
      \path[draw=white,line width=0.608841bp,fill=white] (1.56699,4.74010) -- (2.65576,7.16735) -- (2.73254,5.90566) -- (3.99422,5.82888) -- cycle;
      \path[draw=black,line width=0.797010bp] (27.34605,30.51919) -- (2.90340,6.07653);
      \path[draw=black,line width=0.563565bp,fill=black] (1.50253,4.67565) -- (2.67801,7.29125) -- (2.76248,5.93560) -- (4.11813,5.85113) -- cycle;
      \node[anchor=base west,inner sep=0pt,outer sep=0pt] at (11.14017,9.15285) {\fontsize{10.95pt}{13.6pt}\selectfont $#1$};
      \node[anchor=base west,inner sep=0pt,outer sep=0pt] at (11.68216,31.37585) {\fontsize{10.95pt}{13.6pt}\selectfont $#2$};
      \pgfresetboundingbox
      \path[use as bounding box] (0,0) rectangle (28.04605,34.10311);
    \end{tikzpicture}%
  }}}%
}

\newcommand{\diagWmate}[2]{%
  \mathord{\vcenter{\hbox{%
    \begin{tikzpicture}[x=1bp,y=-1bp,line cap=butt,line join=miter]
      \path[draw=black,line width=0.797010bp] (55.01053,5.69594) .. controls (55.01053,21.68352) and (55.01053,48.32958) .. (41.68750,35.00655) -- (15.04144,8.36049) .. controls (1.71841,-4.96253) and (1.71841,21.68352) .. (1.71841,34.55501);
      \path[draw=black,line width=0.797010bp,fill=black] (1.71835,36.53615) -- (2.73669,33.85546) -- (1.71835,34.75432) -- (0.70001,33.85546) -- cycle;
      \path[draw=black,line width=0.797010bp] (55.01053,5.69594) -- (55.01053,14.57055);
      \path[draw=black,line width=0.797010bp,fill=black] (55.01047,16.55166) -- (56.02881,13.87096) -- (55.01047,14.76983) -- (53.99213,13.87096) -- cycle;
      \path[draw=white,line width=3.586600bp] (15.04144,35.00655) -- (39.50275,10.54525);
      \path[draw=white,line width=0.608841bp,fill=white] (40.82040,9.22759) -- (38.39316,10.31637) -- (39.65485,10.39315) -- (39.73163,11.65484) -- cycle;
      \path[draw=black,line width=0.797010bp] (15.04144,35.00655) -- (39.48410,10.56389);
      \path[draw=black,line width=0.563565bp,fill=black] (40.88486,9.16314) -- (38.26926,10.33861) -- (39.62491,10.42309) -- (39.70938,11.77874) -- cycle;
      \node[anchor=base west,inner sep=0pt,outer sep=0pt] at (25.48169,12.04131) {\fontsize{10.95pt}{13.6pt}\selectfont $#1$};
      \node[anchor=base west,inner sep=0pt,outer sep=0pt] at (26.02371,37.46132) {\fontsize{10.95pt}{13.6pt}\selectfont $#2$};
      \pgfresetboundingbox
      \path[use as bounding box] (0,0) rectangle (56.72882,43.54806);
    \end{tikzpicture}%
  }}}%
}

\newcommand{\diagWtilde}[2]{%
  \mathord{\vcenter{\hbox{%
    \begin{tikzpicture}[x=1bp,y=-1bp,line cap=butt,line join=miter]
      \path[draw=black,line width=0.797010bp] (2.42203,25.43163) -- (26.86469,0.98897);
      \path[draw=black,line width=0.563565bp,fill=black] (1.02127,26.83251) -- (3.63687,25.65703) -- (2.28122,25.57256) -- (2.19675,24.21691) -- cycle;
      \path[draw=white,line width=3.586600bp] (26.86469,27.63503) -- (2.40341,3.17373);
      \path[draw=white,line width=0.608841bp,fill=white] (1.08572,1.85607) -- (2.17450,4.28332) -- (2.25128,3.02163) -- (3.51297,2.94484) -- cycle;
      \path[draw=black,line width=0.797010bp] (26.86469,27.63503) -- (2.42203,3.19237);
      \path[draw=black,line width=0.563565bp,fill=black] (1.02127,1.79162) -- (2.19675,4.40722) -- (2.28122,3.05156) -- (3.63687,2.96709) -- cycle;
      \node[anchor=base west,inner sep=0pt,outer sep=0pt] at (0.00000,16.65981) {\fontsize{10.95pt}{13.6pt}\selectfont $#1$};
      \node[anchor=base west,inner sep=0pt,outer sep=0pt] at (21.85901,18.09981) {\fontsize{10.95pt}{13.6pt}\selectfont $#2$};
      \pgfresetboundingbox
      \path[use as bounding box] (0,0) rectangle (27.56470,28.33504);
    \end{tikzpicture}%
  }}}%
}

\newcommand{\diagV}[2]{%
  \mathord{\vcenter{\hbox{%
    \begin{tikzpicture}[x=1bp,y=-1bp,line cap=butt,line join=miter]
      \path[draw=black,line width=0.797010bp] (27.34607,27.63491) -- (2.90341,3.19222);
      \path[draw=black,line width=0.563565bp,fill=black] (1.50266,1.79140) -- (2.67813,4.40700) -- (2.76260,3.05135) -- (4.11826,2.96688) -- cycle;
      \path[draw=white,line width=3.586600bp] (0.70001,27.63491) -- (25.16132,3.17360);
      \path[draw=white,line width=0.608841bp,fill=white] (26.47910,1.85586) -- (24.05185,2.94460) -- (25.31354,3.02138) -- (25.39032,4.28304) -- cycle;
      \path[draw=black,line width=0.797010bp] (0.70001,27.63491) -- (25.14267,3.19222);
      \path[draw=black,line width=0.563565bp,fill=black] (26.54355,1.79140) -- (23.92795,2.96688) -- (25.28360,3.05135) -- (25.36807,4.40700) -- cycle;
      \node[anchor=base west,inner sep=0pt,outer sep=0pt] at (1.02328,18.09975) {\fontsize{10.95pt}{13.6pt}\selectfont $#1$};
      \node[anchor=base west,inner sep=0pt,outer sep=0pt] at (21.79828,16.65975) {\fontsize{10.95pt}{13.6pt}\selectfont $#2$};
      \pgfresetboundingbox
      \path[use as bounding box] (0,0) rectangle (28.04608,28.33492);
    \end{tikzpicture}%
  }}}%
}

\newcommand{\diagVbar}[2]{%
  \mathord{\vcenter{\hbox{%
    \begin{tikzpicture}[x=1bp,y=-1bp,line cap=butt,line join=miter]
      \path[draw=black,line width=0.797010bp] (0.70001,2.70028) -- (25.14267,27.14294);
      \path[draw=black,line width=0.563565bp,fill=black] (26.54352,28.54384) -- (25.36804,25.92825) -- (25.28357,27.28389) -- (23.92792,27.36837) -- cycle;
      \path[draw=white,line width=3.586600bp] (0.70001,29.34634) -- (25.16129,4.88505);
      \path[draw=white,line width=0.608841bp,fill=white] (26.47906,3.56742) -- (24.05182,4.65619) -- (25.31351,4.73296) -- (25.39029,5.99464) -- cycle;
      \path[draw=black,line width=0.797010bp] (0.70001,29.34634) -- (25.14267,4.90368);
      \path[draw=black,line width=0.563565bp,fill=black] (26.54352,3.50297) -- (23.92792,4.67844) -- (25.28357,4.76291) -- (25.36804,6.11856) -- cycle;
      \node[anchor=base west,inner sep=0pt,outer sep=0pt] at (11.68225,9.15283) {\fontsize{10.95pt}{13.6pt}\selectfont $#1$};
      \node[anchor=base west,inner sep=0pt,outer sep=0pt] at (11.13925,29.02984) {\fontsize{10.95pt}{13.6pt}\selectfont $#2$};
      \pgfresetboundingbox
      \path[use as bounding box] (0,0) rectangle (27.24353,31.75712);
    \end{tikzpicture}%
  }}}%
}

\newcommand{\diagVbubble}[3]{%
  \mathord{\vcenter{\hbox{%
    \begin{tikzpicture}[x=1bp,y=-1bp,line cap=butt,line join=miter]
      \path[draw=black,line width=0.797010bp] (20.20229,4.61401) .. controls (-4.17551,21.67860) and (-4.17551,26.55406) .. (17.64958,41.83174);
      \path[draw=black,line width=0.652918bp,fill=black] (19.27242,42.96780) -- (17.66032,40.59628) -- (17.81271,41.94598) -- (16.49237,42.26474) -- cycle;
      \path[draw=white,line width=3.586600bp] (0.70000,43.61865) .. controls (25.07779,26.55406) and (25.07779,21.67860) .. (3.23114,6.38583);
      \path[draw=white,line width=0.705372bp,fill=white] (1.70444,5.31714) -- (3.19843,7.51820) -- (3.05478,6.26241) -- (4.28392,5.96754) -- cycle;
      \path[draw=black,line width=0.797010bp] (0.70000,43.61865) .. controls (25.07779,26.55406) and (25.07779,21.67860) .. (3.25270,6.40092);
      \path[draw=black,line width=0.652918bp,fill=black] (1.62978,5.26486) -- (3.24187,7.63638) -- (3.08948,6.28668) -- (4.40982,5.96793) -- cycle;
      \node[anchor=base west,inner sep=0pt,outer sep=0pt] at (8.11028,27.90381) {\fontsize{10.95pt}{13.6pt}\selectfont $#2$};
      \node[anchor=base west,inner sep=0pt,outer sep=0pt] at (7.56827,8.18182) {\fontsize{10.95pt}{13.6pt}\selectfont $#1$};
      \node[anchor=base west,inner sep=0pt,outer sep=0pt] at (8.09027,44.74780) {\fontsize{10.95pt}{13.6pt}\selectfont $#3$};
      \pgfresetboundingbox
      \path[use as bounding box] (0,0) rectangle (20.90228,48.41508);
    \end{tikzpicture}%
  }}}%
}

\newcommand{\diagId}[2]{%
  \mathord{\vcenter{\hbox{%
    \begin{tikzpicture}[x=1bp,y=-1bp,line cap=butt,line join=miter]
      \path[draw=black,line width=0.797010bp] (1.71826,25.07780) -- (1.71826,3.81607);
      \path[draw=black,line width=0.797010bp,fill=black] (1.71835,1.83496) -- (0.70001,4.51569) -- (1.71835,3.61682) -- (2.73669,4.51569) -- cycle;
      \path[draw=black,line width=0.797010bp] (18.78265,0.70000) -- (18.78265,21.96173);
      \path[draw=black,line width=0.797010bp,fill=black] (18.78265,23.94284) -- (19.80099,21.26212) -- (18.78265,22.16098) -- (17.76431,21.26212) -- cycle;
      \path[draw=black,line width=0.398500bp,dash pattern=on .3985bp off .99628bp,dash phase=0bp] (1.71826,12.88890) -- (18.78265,12.88890);
      \node[anchor=base west,inner sep=0pt,outer sep=0pt] at (7.36777,8.53337) {\fontsize{10.95pt}{13.6pt}\selectfont $#1$};
      \node[anchor=base west,inner sep=0pt,outer sep=0pt] at (7.89078,21.94139) {\fontsize{10.95pt}{13.6pt}\selectfont $#2$};
      \pgfresetboundingbox
      \path[use as bounding box] (0,0) rectangle (20.50101,25.96021);
    \end{tikzpicture}%
  }}}%
}

\newcommand{\diagIdp}[2]{%
  \mathord{\vcenter{\hbox{%
    \begin{tikzpicture}[x=1bp,y=-1bp,line cap=butt,line join=miter]
      \foreach \x in {1.71826,18.78265}{%
        \path[draw=black,line width=0.797010bp]
          (\x,25.07780) -- (\x,3.81607);
        \path[draw=black,line width=0.797010bp,fill=black]
          (\x,1.83496)
          -- ({\x-1.01834},4.51569)
          -- (\x,3.61682)
          -- ({\x+1.01834},4.51569)
          -- cycle;
      }
      \path[draw=black,line width=0.398500bp,
            dash pattern=on .3985bp off .99628bp,dash phase=0bp]
        (1.71826,12.88890) -- (18.78265,12.88890);
      \node[anchor=base west,inner sep=0pt,outer sep=0pt]
        at (7.36777,8.53337)
        {\fontsize{10.95pt}{13.6pt}\selectfont $#1$};
      \node[anchor=base west,inner sep=0pt,outer sep=0pt]
        at (7.89078,21.94139)
        {\fontsize{10.95pt}{13.6pt}\selectfont $#2$};
      \pgfresetboundingbox
      \path[use as bounding box]
        (0,0) rectangle (20.50101,25.96021);
    \end{tikzpicture}%
  }}}%
}

\newcommand{\diagVtilde}[2]{%
  \mathord{\vcenter{\hbox{%
    \begin{tikzpicture}[x=1bp,y=-1bp,line cap=butt,line join=miter]
      \path[draw=black,line width=0.797010bp] (26.54343,4.13920) -- (2.10077,28.58185);
      \path[draw=black,line width=0.563565bp,fill=black] (0.70001,29.98270) -- (3.31561,28.80723) -- (1.95996,28.72275) -- (1.87549,27.36710) -- cycle;
      \path[draw=white,line width=3.586600bp] (26.54343,30.78525) -- (2.08215,6.32396);
      \path[draw=white,line width=0.608841bp,fill=white] (0.76447,5.00626) -- (1.85324,7.43349) -- (1.93002,6.17182) -- (3.19171,6.09504) -- cycle;
      \path[draw=black,line width=0.797010bp] (26.54343,30.78525) -- (2.10077,6.34260);
      \path[draw=black,line width=0.563565bp,fill=black] (0.70001,4.94181) -- (1.87549,7.55741) -- (1.95996,6.20176) -- (3.31561,6.11728) -- cycle;
      \node[anchor=base west,inner sep=0pt,outer sep=0pt] at (10.33765,9.15180) {\fontsize{10.95pt}{13.6pt}\selectfont $#2$};
      \node[anchor=base west,inner sep=0pt,outer sep=0pt] at (10.88065,31.90778) {\fontsize{10.95pt}{13.6pt}\selectfont $#1$};
      \pgfresetboundingbox
      \path[use as bounding box] (0,0) rectangle (27.24344,34.63506);
    \end{tikzpicture}%
  }}}%
}

\newcommand{\diagVmate}[2]{%
  \mathord{\vcenter{\hbox{%
    \begin{tikzpicture}[x=1bp,y=-1bp,line cap=butt,line join=miter]
      \path[draw=black,line width=0.797010bp] (55.01053,5.69625) .. controls (55.01053,21.68383) and (55.01053,48.32989) .. (41.68750,35.00686) -- (15.04144,8.36080) .. controls (1.71841,-4.96223) and (1.71841,21.68383) .. (1.71841,34.55533);
      \path[draw=black,line width=0.797010bp,fill=black] (1.71835,36.53653) -- (2.73669,33.85581) -- (1.71835,34.75467) -- (0.70001,33.85581) -- cycle;
      \path[draw=black,line width=0.797010bp] (55.01053,5.69625) -- (55.01053,14.57084);
      \path[draw=black,line width=0.797010bp,fill=black] (55.01047,16.55204) -- (56.02881,13.87131) -- (55.01047,14.77018) -- (53.99213,13.87131) -- cycle;
      \path[draw=white,line width=3.586600bp] (15.04144,35.00686) -- (39.50272,10.54557);
      \path[draw=white,line width=0.608841bp,fill=white] (40.82040,9.22793) -- (38.39316,10.31670) -- (39.65485,10.39348) -- (39.73163,11.65516) -- cycle;
      \path[draw=black,line width=0.797010bp] (15.04144,35.00686) -- (39.48410,10.56420);
      \path[draw=black,line width=0.563565bp,fill=black] (40.88486,9.16347) -- (38.26926,10.33895) -- (39.62491,10.42342) -- (39.70938,11.77907) -- cycle;
      \node[anchor=base west,inner sep=0pt,outer sep=0pt] at (15.36569,25.47143) {\fontsize{10.95pt}{13.6pt}\selectfont $#1$};
      \node[anchor=base west,inner sep=0pt,outer sep=0pt] at (36.13971,24.03148) {\fontsize{10.95pt}{13.6pt}\selectfont $#2$};
      \pgfresetboundingbox
      \path[use as bounding box] (0,0) rectangle (56.72882,43.54836);
    \end{tikzpicture}%
  }}}%
}

\newcommand{\diagVinv}[2]{%
  \mathord{\vcenter{\hbox{%
    \begin{tikzpicture}[x=1bp,y=-1bp,line cap=butt,line join=miter]
      \path[draw=black,line width=0.797010bp] (0.70001,27.63320) -- (25.14267,3.19054);
      \path[draw=black,line width=0.563565bp,fill=black] (26.54352,1.78979) -- (23.92792,2.96526) -- (25.28357,3.04973) -- (25.36804,4.40539) -- cycle;
      \path[draw=white,line width=3.586600bp] (27.34607,27.63320) -- (2.88477,3.17193);
      \path[draw=white,line width=0.608841bp,fill=white] (1.56702,1.85424) -- (2.65579,4.28149) -- (2.73257,3.01980) -- (3.99426,2.94301) -- cycle;
      \path[draw=black,line width=0.797010bp] (27.34607,27.63320) -- (2.90341,3.19054);
      \path[draw=black,line width=0.563565bp,fill=black] (1.50256,1.78979) -- (2.67804,4.40539) -- (2.76251,3.04973) -- (4.11816,2.96526) -- cycle;
      \node[anchor=base west,inner sep=0pt,outer sep=0pt] at (1.02328,18.09786) {\fontsize{10.95pt}{13.6pt}\selectfont $#1$};
      \node[anchor=base west,inner sep=0pt,outer sep=0pt] at (21.79828,16.65887) {\fontsize{10.95pt}{13.6pt}\selectfont $#2$};
      \pgfresetboundingbox
      \path[use as bounding box] (0,0) rectangle (28.04608,28.33321);
    \end{tikzpicture}%
  }}}%
}

\newcommand{\diagVinvbar}[2]{%
  \mathord{\vcenter{\hbox{%
    \begin{tikzpicture}[x=1bp,y=-1bp,line cap=butt,line join=miter]
      \path[draw=black,line width=0.797010bp] (0.70001,29.34515) -- (25.14267,4.90249);
      \path[draw=black,line width=0.563565bp,fill=black] (26.54352,3.50164) -- (23.92792,4.67712) -- (25.28357,4.76159) -- (25.36804,6.11724) -- cycle;
      \path[draw=white,line width=3.586600bp] (0.70001,2.69909) -- (25.16129,27.16039);
      \path[draw=white,line width=0.608841bp,fill=white] (26.47906,28.47817) -- (25.39029,26.05093) -- (25.31351,27.31261) -- (24.05182,27.38940) -- cycle;
      \path[draw=black,line width=0.797010bp] (0.70001,2.69909) -- (25.14267,27.14175);
      \path[draw=black,line width=0.563565bp,fill=black] (26.54352,28.54262) -- (25.36804,25.92702) -- (25.28357,27.28268) -- (23.92792,27.36715) -- cycle;
      \node[anchor=base west,inner sep=0pt,outer sep=0pt] at (11.68225,9.15084) {\fontsize{10.95pt}{13.6pt}\selectfont $#1$};
      \node[anchor=base west,inner sep=0pt,outer sep=0pt] at (11.14026,29.02782) {\fontsize{10.95pt}{13.6pt}\selectfont $#2$};
      \pgfresetboundingbox
      \path[use as bounding box] (0,0) rectangle (27.24353,31.75509);
    \end{tikzpicture}%
  }}}%
}

\newcommand{\diagVinvbubble}[3]{%
  \mathord{\vcenter{\hbox{%
    \begin{tikzpicture}[x=1bp,y=-1bp,line cap=butt,line join=miter]
      \path[draw=black,line width=0.797010bp] (0.70000,43.61945) .. controls (25.07779,26.55487) and (25.07779,21.67938) .. (3.25270,6.40170);
      \path[draw=black,line width=0.652918bp,fill=black] (1.62976,5.26566) -- (3.24185,7.63718) -- (3.08946,6.28748) -- (4.40981,5.96872) -- cycle;
      \path[draw=white,line width=3.586600bp] (20.20229,4.61481) .. controls (-4.17551,21.67938) and (-4.17551,26.55487) .. (17.67114,41.84763);
      \path[draw=white,line width=0.705372bp,fill=white] (19.19774,42.91632) -- (17.70375,40.71524) -- (17.84740,41.97107) -- (16.61826,42.26590) -- cycle;
      \path[draw=black,line width=0.797010bp] (20.20229,4.61481) .. controls (-4.17551,21.67938) and (-4.17551,26.55487) .. (17.64958,41.83255);
      \path[draw=black,line width=0.652918bp,fill=black] (19.27240,42.96860) -- (17.66031,40.59708) -- (17.81270,41.94678) -- (16.49236,42.26553) -- cycle;
      \node[anchor=base west,inner sep=0pt,outer sep=0pt] at (8.11026,27.90482) {\fontsize{10.95pt}{13.6pt}\selectfont $#2$};
      \node[anchor=base west,inner sep=0pt,outer sep=0pt] at (7.56827,8.18182) {\fontsize{10.95pt}{13.6pt}\selectfont $#1$};
      \node[anchor=base west,inner sep=0pt,outer sep=0pt] at (8.09126,44.74884) {\fontsize{10.95pt}{13.6pt}\selectfont $#3$};
      \pgfresetboundingbox
      \path[use as bounding box] (0,0) rectangle (20.90228,48.41514);
    \end{tikzpicture}%
  }}}%
}

\newcommand{\diagStar}[4]{%
  \mathord{\vcenter{\hbox{%
    \begin{tikzpicture}[x=1bp,y=-1bp,line cap=butt,line join=miter]
      \path[draw=black,line width=0.797010bp] (43.46173,50.89105) -- (2.25774,18.51636);
      \path[draw=black,line width=0.626681bp,fill=black] (0.70000,17.29254) -- (2.17870,19.74939) -- (2.10104,18.39337) -- (3.43695,18.14801) -- cycle;
      \path[draw=white,line width=3.586600bp] (32.54819,54.00916) -- (32.54819,10.32642);
      \path[draw=white,line width=0.861040bp,fill=white] (32.54816,8.46283) -- (31.60172,10.94904) -- (32.54816,10.11115) -- (33.49460,10.94904) -- cycle;
      \path[draw=black,line width=0.797010bp] (32.54819,54.00916) -- (32.54819,10.35278);
      \path[draw=black,line width=0.797010bp,fill=black] (32.54816,8.37164) -- (31.52982,11.05237) -- (32.54816,10.15350) -- (33.56650,11.05237) -- cycle;
      \path[draw=white,line width=3.586600bp] (2.92575,50.89105) -- (41.35944,9.50079);
      \path[draw=white,line width=0.585868bp,fill=white] (42.62747,8.13519) -- (40.24228,9.31305) -- (41.50589,9.34308) -- (41.62933,10.60101) -- cycle;
      \path[draw=black,line width=0.797010bp] (2.92575,50.89105) -- (41.34149,9.52008);
      \path[draw=black,line width=0.542301bp,fill=black] (42.68948,8.06842) -- (40.11926,9.33990) -- (41.47708,9.37408) -- (41.61166,10.72565) -- cycle;
      \node[anchor=base west,inner sep=0pt,outer sep=0pt] at (34.34189,8.18182) {\fontsize{10.95pt}{13.6pt}\selectfont $#1$};
      \node[anchor=base west,inner sep=0pt,outer sep=0pt] at (34.88391,56.23682) {\fontsize{10.95pt}{13.6pt}\selectfont $#2$};
      \node[anchor=base west,inner sep=0pt,outer sep=0pt] at (6.80191,34.52985) {\fontsize{10.95pt}{13.6pt}\selectfont $#3$};
      \node[anchor=base west,inner sep=0pt,outer sep=0pt] at (25.03189,34.41083) {\fontsize{10.95pt}{13.6pt}\selectfont $#4$};
      \pgfresetboundingbox
      \path[use as bounding box] (0,0) rectangle (44.16174,63.30509);
    \end{tikzpicture}%
  }}}%
}

\newcommand{\diagTriangle}[3]{%
  \mathord{\vcenter{\hbox{%
    \begin{tikzpicture}[x=1bp,y=-1bp,line cap=butt,line join=miter]
      \path[draw=black,line width=0.797010bp] (44.35413,36.98291) -- (3.15012,4.60828);
      \path[draw=black,line width=0.626681bp,fill=black] (1.59238,3.38440) -- (3.07108,5.84125) -- (2.99344,4.48523) -- (4.32932,4.23987) -- cycle;
      \path[draw=white,line width=3.586600bp] (11.61356,46.33752) -- (11.61356,2.65479);
      \path[draw=white,line width=0.861040bp,fill=white] (11.61356,0.79120) -- (10.66711,3.27740) -- (11.61356,2.43951) -- (12.56000,3.27740) -- cycle;
      \path[draw=black,line width=0.797010bp] (11.61356,46.33752) -- (11.61356,2.68115);
      \path[draw=black,line width=0.797010bp,fill=black] (11.61356,0.70001) -- (10.59521,3.38074) -- (11.61356,2.48187) -- (12.63190,3.38074) -- cycle;
      \path[draw=white,line width=3.586600bp] (0.70001,46.33752) -- (42.16937,4.86798);
      \path[draw=white,line width=0.608841bp,fill=white] (43.48709,3.55029) -- (41.05984,4.63904) -- (42.32153,4.71582) -- (42.39832,5.97748) -- cycle;
      \path[draw=black,line width=0.797010bp] (0.70001,46.33752) -- (42.15073,4.88660);
      \path[draw=black,line width=0.563565bp,fill=black] (43.55154,3.48584) -- (40.93594,4.66132) -- (42.29160,4.74579) -- (42.37607,6.10144) -- cycle;
      \node[anchor=base west,inner sep=0pt,outer sep=0pt] at (22.76230,12.82721) {\fontsize{10.95pt}{13.6pt}\selectfont $#1$};
      \node[anchor=base west,inner sep=0pt,outer sep=0pt] at (23.30429,36.87219) {\fontsize{10.95pt}{13.6pt}\selectfont $#2$};
      \node[anchor=base west,inner sep=0pt,outer sep=0pt] at (2.23828,26.85822) {\fontsize{10.95pt}{13.6pt}\selectfont $#3$};
      \pgfresetboundingbox
      \path[use as bounding box] (0,0) rectangle (45.05414,47.96183);
    \end{tikzpicture}%
  }}}%
}

\newcommand{\diagWbar}[2]{%
  \mathord{\vcenter{\hbox{%
    \begin{tikzpicture}[x=1bp,y=-1bp,line cap=butt,line join=miter]
      \path[draw=black,line width=0.797010bp] (0.70001,1.79139) -- (25.14267,26.23404);
      \path[draw=black,line width=0.563565bp,fill=black] (26.54352,27.63495) -- (25.36804,25.01935) -- (25.28357,26.37500) -- (23.92792,26.45947) -- cycle;
      \path[draw=white,line width=3.586600bp] (0.70001,28.43744) -- (25.16129,3.97616);
      \path[draw=white,line width=0.608841bp,fill=white] (26.47906,2.65852) -- (24.05182,3.74729) -- (25.31351,3.82406) -- (25.39029,5.08574) -- cycle;
      \path[draw=black,line width=0.797010bp] (0.70001,28.43744) -- (25.14267,3.99479);
      \path[draw=black,line width=0.563565bp,fill=black] (26.54352,2.59407) -- (23.92792,3.76954) -- (25.28357,3.85402) -- (25.36804,5.20966) -- cycle;
      \node[anchor=base west,inner sep=0pt,outer sep=0pt] at (0.62201,18.50101) {\fontsize{10.95pt}{13.6pt}\selectfont $#2$};
      \node[anchor=base west,inner sep=0pt,outer sep=0pt] at (21.39700,17.06101) {\fontsize{10.95pt}{13.6pt}\selectfont $#1$};
      \pgfresetboundingbox
      \path[use as bounding box] (0,0) rectangle (27.24353,29.13744);
    \end{tikzpicture}%
  }}}%
}

\newcommand{\diagWminusbar}[2]{%
  \mathord{\vcenter{\hbox{%
    \begin{tikzpicture}[x=1bp,y=-1bp,line cap=butt,line join=miter]
      \path[draw=black,line width=0.797010bp] (0.70001,28.43746) -- (25.14267,3.99480);
      \path[draw=black,line width=0.563565bp,fill=black] (26.54352,2.59395) -- (23.92792,3.76943) -- (25.28357,3.85390) -- (25.36804,5.20955) -- cycle;
      \path[draw=white,line width=3.586600bp] (0.70001,1.79140) -- (25.16129,26.25271);
      \path[draw=white,line width=0.608841bp,fill=white] (26.47906,27.57049) -- (25.39029,25.14324) -- (25.31351,26.40493) -- (24.05182,26.48171) -- cycle;
      \path[draw=black,line width=0.797010bp] (0.70001,1.79140) -- (25.14267,26.23406);
      \path[draw=black,line width=0.563565bp,fill=black] (26.54352,27.63494) -- (25.36804,25.01934) -- (25.28357,26.37499) -- (23.92792,26.45946) -- cycle;
      \node[anchor=base west,inner sep=0pt,outer sep=0pt] at (0.62201,18.50103) {\fontsize{10.95pt}{13.6pt}\selectfont $#2$};
      \node[anchor=base west,inner sep=0pt,outer sep=0pt] at (21.39700,17.06102) {\fontsize{10.95pt}{13.6pt}\selectfont $#1$};
      \pgfresetboundingbox
      \path[use as bounding box] (0,0) rectangle (27.24353,29.13747);
    \end{tikzpicture}%
  }}}%
}

\newcommand{\diagWbubble}[3]{%
  \mathord{\vcenter{\hbox{%
    \begin{tikzpicture}[x=1bp,y=-1bp,line cap=butt,line join=miter]
      \path[draw=black,line width=0.797010bp] (0.70000,43.61945) .. controls (25.07779,26.55487) and (25.07779,21.67938) .. (3.25270,6.40170);
      \path[draw=black,line width=0.652918bp,fill=black] (1.62976,5.26566) -- (3.24185,7.63718) -- (3.08946,6.28748) -- (4.40981,5.96872) -- cycle;
      \path[draw=white,line width=3.586600bp] (20.20229,43.61945) .. controls (-4.17551,26.55487) and (-4.17551,21.67938) .. (17.64958,6.40170);
      \path[draw=white,line width=0.705152bp,fill=white] (19.27252,5.26566) -- (17.66043,7.63718) -- (17.81282,6.28748) -- (16.49248,5.96872) -- cycle;
      \path[draw=black,line width=0.797010bp] (20.20229,43.61945) .. controls (-4.17551,26.55487) and (-4.17551,21.67938) .. (17.64958,6.40170);
      \path[draw=black,line width=0.652918bp,fill=black] (19.27252,5.26566) -- (17.66043,7.63718) -- (17.81282,6.28748) -- (16.49248,5.96872) -- cycle;
      \node[anchor=base west,inner sep=0pt,outer sep=0pt] at (8.11026,27.90482) {\fontsize{10.95pt}{13.6pt}\selectfont $#2$};
      \node[anchor=base west,inner sep=0pt,outer sep=0pt] at (7.56827,8.18182) {\fontsize{10.95pt}{13.6pt}\selectfont $#1$};
      \node[anchor=base west,inner sep=0pt,outer sep=0pt] at (8.09126,44.74884) {\fontsize{10.95pt}{13.6pt}\selectfont $#3$};
      \pgfresetboundingbox
      \path[use as bounding box] (0,0) rectangle (20.90228,48.41514);
    \end{tikzpicture}%
  }}}%
}

\newcommand{\diagWminusmate}[2]{%
  \mathord{\vcenter{\hbox{%
    \begin{tikzpicture}[x=1bp,y=-1bp,line cap=butt,line join=miter]
      \path[draw=black,line width=0.797010bp] (15.04144,35.00655) -- (39.48410,10.56389);
      \path[draw=black,line width=0.563565bp,fill=black] (40.88486,9.16314) -- (38.26926,10.33861) -- (39.62491,10.42309) -- (39.70938,11.77874) -- cycle;
      \path[draw=white,line width=3.586600bp] (55.01053,5.69594) .. controls (55.01053,21.68352) and (55.01053,48.32958) .. (41.68750,35.00655) -- (15.04144,8.36049) .. controls (1.71841,-4.96253) and (1.71841,21.68352) .. (1.71841,34.55501);
      \path[draw=black,line width=0.797010bp] (55.01053,5.69594) .. controls (55.01053,21.68352) and (55.01053,48.32958) .. (41.68750,35.00655) -- (15.04144,8.36049) .. controls (1.71841,-4.96253) and (1.71841,21.68352) .. (1.71841,34.55501);
      \path[draw=black,line width=0.797010bp,fill=black] (1.71835,36.53615) -- (2.73669,33.85546) -- (1.71835,34.75432) -- (0.70001,33.85546) -- cycle;
      \path[draw=black,line width=0.797010bp] (55.01053,5.69594) -- (55.01053,14.57055);
      \path[draw=black,line width=0.797010bp,fill=black] (55.01047,16.55166) -- (56.02881,13.87096) -- (55.01047,14.76983) -- (53.99213,13.87096) -- cycle;
      \node[anchor=base west,inner sep=0pt,outer sep=0pt] at (25.48169,12.04131) {\fontsize{10.95pt}{13.6pt}\selectfont $#1$};
      \node[anchor=base west,inner sep=0pt,outer sep=0pt] at (26.02371,37.46132) {\fontsize{10.95pt}{13.6pt}\selectfont $#2$};
      \pgfresetboundingbox
      \path[use as bounding box] (0,0) rectangle (56.72882,43.54806);
    \end{tikzpicture}%
  }}}%
}

\newcommand{\diagVminusmate}[2]{%
  \mathord{\vcenter{\hbox{%
    \begin{tikzpicture}[x=1bp,y=-1bp,line cap=butt,line join=miter]
      \path[draw=black,line width=0.797010bp] (15.04144,35.00655) -- (39.48410,10.56389);
      \path[draw=black,line width=0.563565bp,fill=black] (40.88486,9.16314) -- (38.26926,10.33861) -- (39.62491,10.42309) -- (39.70938,11.77874) -- cycle;
      \path[draw=white,line width=3.586600bp] (55.01053,5.69594) .. controls (55.01053,21.68352) and (55.01053,48.32958) .. (41.68750,35.00655) -- (15.04144,8.36049) .. controls (1.71841,-4.96253) and (1.71841,21.68352) .. (1.71841,34.55501);
      \path[draw=black,line width=0.797010bp] (55.01053,5.69594) .. controls (55.01053,21.68352) and (55.01053,48.32958) .. (41.68750,35.00655) -- (15.04144,8.36049) .. controls (1.71841,-4.96253) and (1.71841,21.68352) .. (1.71841,34.55501);
      \path[draw=black,line width=0.797010bp,fill=black] (1.71835,36.53615) -- (2.73669,33.85546) -- (1.71835,34.75432) -- (0.70001,33.85546) -- cycle;
      \path[draw=black,line width=0.797010bp] (55.01053,5.69594) -- (55.01053,14.57055);
      \path[draw=black,line width=0.797010bp,fill=black] (55.01047,16.55166) -- (56.02881,13.87096) -- (55.01047,14.76983) -- (53.99213,13.87096) -- cycle;
      \node[anchor=base west,inner sep=0pt,outer sep=0pt] at (35,24) {\fontsize{10.95pt}{13.6pt}\selectfont $#1$};
      \node[anchor=base west,inner sep=0pt,outer sep=0pt] at (16,24) {\fontsize{10.95pt}{13.6pt}\selectfont $#2$};
      \pgfresetboundingbox
      \path[use as bounding box] (0,0) rectangle (56.72882,43.54806);
    \end{tikzpicture}%
  }}}%
}

\newcommand{\diagWminustilde}[2]{%
  \mathord{\vcenter{\hbox{%
    \begin{tikzpicture}[x=1bp,y=-1bp,line cap=butt,line join=miter]
      \path[draw=black,line width=0.797010bp] (26.86469,27.63503) -- (2.42203,3.19237);
      \path[draw=black,line width=0.563565bp,fill=black] (1.02127,1.79162) -- (2.19675,4.40722) -- (2.28122,3.05156) -- (3.63687,2.96709) -- cycle;
      \path[draw=white,line width=3.586600bp] (2.42203,25.43163) -- (26.86469,0.98897);
      \path[draw=black,line width=0.797010bp] (2.42203,25.43163) -- (26.86469,0.98897);
      \path[draw=black,line width=0.563565bp,fill=black] (1.02127,26.83251) -- (3.63687,25.65703) -- (2.28122,25.57256) -- (2.19675,24.21691) -- cycle;
      \node[anchor=base west,inner sep=0pt,outer sep=0pt] at (0.00000,16.65981) {\fontsize{10.95pt}{13.6pt}\selectfont $#1$};
      \node[anchor=base west,inner sep=0pt,outer sep=0pt] at (21.85901,18.09981) {\fontsize{10.95pt}{13.6pt}\selectfont $#2$};
      \pgfresetboundingbox
      \path[use as bounding box] (0,0) rectangle (27.56470,28.33504);
    \end{tikzpicture}%
  }}}%
}

\newcommand{\diagVminustilde}[2]{%
  \mathord{\vcenter{\hbox{%
    \begin{tikzpicture}[x=1bp,y=-1bp,line cap=butt,line join=miter]
      \path[draw=black,line width=0.797010bp] (26.86469,27.63503) -- (2.42203,3.19237);
      \path[draw=black,line width=0.563565bp,fill=black] (1.02127,1.79162) -- (2.19675,4.40722) -- (2.28122,3.05156) -- (3.63687,2.96709) -- cycle;
      \path[draw=white,line width=3.586600bp] (2.42203,25.43163) -- (26.86469,0.98897);
      \path[draw=black,line width=0.797010bp] (2.42203,25.43163) -- (26.86469,0.98897);
      \path[draw=black,line width=0.563565bp,fill=black] (1.02127,26.83251) -- (3.63687,25.65703) -- (2.28122,25.57256) -- (2.19675,24.21691) -- cycle;
      \node[anchor=base west,inner sep=0pt,outer sep=0pt] at (10.00000,7.65981) {\fontsize{10.95pt}{13.6pt}\selectfont $#1$};
      \node[anchor=base west,inner sep=0pt,outer sep=0pt] at (11,26.09981) {\fontsize{10.95pt}{13.6pt}\selectfont $#2$};
      \pgfresetboundingbox
      \path[use as bounding box] (0,0) rectangle (27.56470,28.33504);
    \end{tikzpicture}%
  }}}%
}
\newcommand{\diagEvCo}[3]{%
  \mathord{\vcenter{\hbox{%
    \begin{tikzpicture}[x=1bp,y=#2bp,line cap=round,line join=miter]
      \path[draw=black,line width=0.797010bp]
        (28,0) .. controls (28,10) and (22,18) .. (14,18)
               .. controls (6,18) and (0,10) .. (0,0);
      \path[draw=black,line width=0.797010bp,fill=black]
        (12.5,18) -- (15.18073,19.01834) -- (14.28186,18)
        -- (15.18073,16.98166) -- cycle;
      \node[inner sep=0pt,outer sep=0pt] at (14,#3) {$#1$};
      \path[use as bounding box] (-1,-14) rectangle (29,34);
    \end{tikzpicture}%
  }}}%
}
\newcommand{\diagEv}[1]{\diagEvCo{#1}{1}{7}}
\newcommand{\diagEvOutside}[1]{\diagEvCo{#1}{1}{24}}
\newcommand{\diagCoev}[1]{\diagEvCo{#1}{-1}{7}}
\newcommand{\diagCoevOutside}[1]{\diagEvCo{#1}{-1}{24}}

\newtheorem{proposition}{Proposition}
\newtheorem{lemma}{Lemma}
\theoremstyle{remark}
\newtheorem{definition}{Definition}
\newtheorem{remark}{Remark}
\newtheorem{example}{Example}

\usepackage[colorlinks=true,
            linkcolor=blue,
            citecolor=blue,
            urlcolor=blue]{hyperref}
            
\begin{document}
\begin{abstract}
 We propose a matrix generalization of the star--triangle relations and identify solutions underlying the $\mathfrak{sl}_n$ chiral Potts models.
\end{abstract}
\maketitle
\section{Introduction}

The star--triangle relations (STR) occupy an important place in the theory of exactly solvable models of statistical mechanics and in integrable systems arising in quantum field theory~\cite{MR10315,MR690578,PerkAuYang2006,Spiridonov2012EllipticBeta,MR3019403}. Among their most remarkable solutions are the chiral Potts models of Baxter, Perk, and Au-Yang~\cite{BaxterPerkAuYang1988}, which generalize the self-dual Fateev--Zamolodchikov models~\cite{FateevZamolodchikov1982}. These provided the first examples of solvable lattice models in which the rapidity variables lie on algebraic curves of genus greater than one, in sharp contrast with the rational, trigonometric, and elliptic parametrizations familiar from earlier models.

A further important development was the discovery of the relation between the chiral Potts models, the six-vertex model~\cite{Sutherland1967,Lieb1967}, and the representation theory of the quantum group $U_q(\mathfrak{sl}_2)$ at roots of unity~\cite{BazhanovStroganov1990}. This connection led to a broader representation-theoretic understanding of the model and stimulated higher-rank generalizations associated with quantum groups $U_q(\mathfrak{sl}_n)$~\cite{DJMM1991,BKMS1991,MR1123212,MR1099699,MR1187561}.

These generalized chiral Potts models, however, do not arise directly as solutions of the ordinary STR. Instead, their $R$-matrices satisfy vertex-type Yang--Baxter equations (YBE) of a special factorized form~\cite{DJMM1991,MR1123212}. The corresponding factorization is naturally expressed in terms of so-called star--star relations. Such relations are closely related to Baxter's local integrability condition for three-dimensional lattice models and were developed explicitly in the Baxter--Bazhanov model~\cite{MR1193849}. They subsequently appeared in a number of related works~\cite{MR1210206,KashaevMangazeevStroganov1993Spatial,Baxter1997StarStar,MR2868616,BazhanovKelsSergeev2013}. Thus the YBE, the STR, and the star--star relations are different, although related, manifestations of local integrability.

Constant versions of the STR play an important role in quantum topology, in analogy with applications of solutions of the constant YBE~\cite{AkutsuWadati1987,Turaev1988,MR1025161,MR1036112}. A particularly influential observation, going back to Jones~\cite{MR990215}, is that statistical-mechanical spin models satisfying suitable constant STR can be used to construct invariants of knots and links. This point of view was developed systematically in the theory of spin models and generalized spin models, notably in work of Jaeger~\cite{jaeger1992strongly,Jaeger1993Plane,Jaeger1995SpinDuality,Jaeger1995Survey,MR1433983,Jaeger2000FourWeight}, de la Harpe~\cite{deLaHarpe1994}, Kawagoe--Munemasa--Watatani~\cite{KawagoeMunemasaWatatani1994}, Nomura~\cite{Nomura1995Twisted}, Bannai--Jaeger--Sali~\cite{BannaiJaegerSali1994}, and others. In these constructions, the constant STR becomes an algebraic identity ensuring invariance of the associated partition function under local transformations of link diagrams.

A direct passage from a rapidity-dependent STR solution to a constant spin-model solution used in link invariants was exhibited by Wu, Pant, and King for the self-dual chiral Potts model by taking an infinite-rapidity limit of the Boltzmann weights~\cite{WuPantKing1994,WuPantKing1995}. 

In this work, we propose a matrix generalization of the STR that contains the usual STR and the YBE as special cases. We give explicit examples of both constant and rapidity-dependent solutions that correspond to neither the STR nor the YBE. We then identify the solutions underlying the $\mathfrak{sl}_n$ chiral Potts models, which, for $n\geq 3$, to the best of our knowledge, have been missing from the literature.

\section{Definition of the matrix star--triangle relations and first examples}
For a $\Bbbk$-vector space $a$, we use evaluation and coevaluation maps, represented graphically by caps and cups oriented from right to left:
\[
 \ev_a=\diagEv{a}=\diagEvOutside{a}
 :a^*\otimes a\longrightarrow\Bbbk,\qquad
 \coev_a=\diagCoev{a}=\diagCoevOutside{a}
 :\Bbbk\longrightarrow a\otimes a^*,
\]
where, depending on the context, the vector-space label is associated with one of the two adjacent regions (but not both), so it may be placed on either side of the curve without changing the algebraic meaning of the map.

Let $A$ be a finite set of finite-dimensional $\Bbbk$-vector spaces.
Consider linear maps
\[
 w_{a,b}=\diagW{a}{b},\quad \dw_{a,b}=\diagWminus{a}{b}:b\otimes b\longrightarrow a\otimes a,\quad a,b\in A,
\]
satisfying the identities
\begin{equation}
 \sum_{b\in A}\diagWbubble{a}{b}{c}
 =\diagIdp{a}{c}
 \quad\Longleftrightarrow\quad
 \sum_{b\in A}w_{a,b}\dw_{b,c}=\delta_{a,c}\,\id_{a\otimes a}, 
\end{equation}
and admitting the inverses
\[
 \overline w_{a,b}=(\widetilde w_{b,a})^{-1}
 =\diagWbar{a}{b},\quad  \overline{\dw}_{a,b}=\big(\widetilde{\dw}_{b,a}\big)^{-1}
 =\diagWminusbar{a}{b}:b\otimes a^*\longrightarrow b^*\otimes a,
\]
where
\[
 \widetilde w_{a,b}=\diagWtilde{a}{b}=\diagWmate{a}{b}, \quad  \widetilde{\dw}_{a,b}=\diagWminustilde{a}{b}=\diagWminusmate{a}{b}
\]
or, algebraically,
\begin{equation}
 \widetilde w_{a,b}=(\ev_a\otimes\id_{a\otimes b^*})
 (\id_{a^*}\otimes w_{a,b}\otimes\id_{b^*})(\id_{a^*\otimes b}\otimes\coev_b)
 \colon a^*\otimes b\longrightarrow a\otimes b^*,
\end{equation}
and similarly for $\dw_{a,b}$.

Let $B$ be another finite set of finite-dimensional $\Bbbk$-vector spaces such that
\begin{equation}
\sum_{b\in B}(\dim b)^2= \sum_{a\in A}(\dim a)^2.
\end{equation}
For any $a\in A$ and $b\in B$, we consider invertible linear maps
\[
 v_{b,a}=\diagV{b}{a},\quad  v^{-1}_{b,a}=\diagVinv{b}{a}:b\otimes a\longrightarrow b\otimes a,
\]
for which there exist linear maps
\[
 \overline v_{b,a}=\diagVbar{b}{a},\quad  \overline{v_{b,a}^{-1}}=\diagVinvbar{b}{a} :a\otimes a^*\longrightarrow b^*\otimes b
\]
such that, for any $a,c\in A$,
\begin{equation}
 \sum_{b\in B}\diagVbubble{a}{b}{c}=\diagId{a}{c}\quad \Longleftrightarrow\quad
 \sum_{b\in B}\widetilde v_{b,a}\overline v_{b,c}
 =\delta_{a,c}\,\id_{a\otimes a^*}
\end{equation}
and
\begin{equation}
 \sum_{b\in B}\diagVinvbubble{a}{b}{c}=\diagId{a}{c}\quad\Longleftrightarrow\quad
 \sum_{b\in B}\widetilde{v_{b,a}^{-1}}\,\overline{v_{b,c}^{-1}}
 =\delta_{a,c}\,\id_{a\otimes a^*},
\end{equation}
where
\[
 \widetilde v_{b,a}=\diagVtilde{b}{a}=\diagVmate{b}{a},\quad \widetilde{v_{b,a}^{-1}}=\diagVminustilde{a}{b}=\diagVminusmate{a}{b}
\]
or, algebraically,
\begin{equation}
 \widetilde v_{b,a}=(\ev_b\otimes\id_{a\otimes a^*})
 (\id_{b^*}\otimes v_{b,a}\otimes\id_{a^*})(\id_{b^*\otimes b}\otimes\coev_a)
 \colon b^*\otimes b\longrightarrow a\otimes a^*
\end{equation}
and similarly for $v_{b,a}^{-1}.$

\begin{definition}\label{def:mstr}
 Let $A,B$ and the families
$W=\{w_{a,b}\}_{a,b\in A}$ and
$V=\{v_{b,a}\}_{a\in A,\,b\in B}$ be as above.
We say that $W$ and $V$ satisfy the \emph{Matrix Star-Triangle Relations (MSTR)} if
\begin{equation}\label{eq:mstr}
 \sum_{d\in A}\diagStar{a}{c}{b}{d}=\diagTriangle{a}{c}{b},
\end{equation}
that is,
\begin{equation}\label{eq:mstr-alg}
\sum_{d\in A}(\id_b\otimes w_{a,d})(v_{b,d}\otimes\id_d)(\id_b\otimes w_{d,c})=(v_{b,a}\otimes\id_a)(\id_b\otimes w_{a,c})(v_{b,c}\otimes\id_c)
\end{equation}
for all $a,c\in A$ and $b\in B$.\hfill$\diamond$
\end{definition}
\begin{remark}
 The main idea behind Definition~\ref{def:mstr} is that, in the graphical notation~\eqref{eq:mstr}, every oriented segment adjacent to a region labelled by a vector space $a$ corresponds either to the vector space $a$ or its dual $a^*$, depending on the projection of the segment onto the vertical axis.\hfill$\diamond$
\end{remark}
\begin{remark}
 Definition~\ref{def:mstr} gives the constant MSTR. The rapidity-dependent version is obtained by assigning a rapidity variable to each straight line in~\eqref{eq:mstr} and allowing the maps associated with crossings to depend on the corresponding rapidities. Concretely, we assign the rapidities \(p,q,r\) to the lines with zero, one, and two breaks, respectively. We refer to these as the uppermost, middle, and lowermost lines. The resulting rapidity-dependent MSTR take the form
 \begin{multline}\label{eq:MSTR-rapidity}
\sum_{d\in A}(\id_b\otimes w_{a,d}(p,q))(v_{b,d}(p,r)\otimes\id_d)(\id_b\otimes w_{d,c}(q,r))\\
=(v_{b,a}(q,r)\otimes\id_a)(\id_b\otimes w_{a,c}(p,r))(v_{b,c}(p,q)\otimes\id_c).
\end{multline}
\hfill$\diamond$
\end{remark}
\begin{remark}
The MSTR simultaneously generalize both the STR and the (usual vertex-type) YBE: the STR correspond to the case in which all vector spaces in $A$ and $B$ are one-dimensional, while the YBE is a special case of the situation in which both $A$ and $B$ are singletons. More generally, this latter singleton case corresponds to the generalized YBE of~\cite{MR1234169}.\hfill$\diamond$
\end{remark}
\begin{remark}
 Solutions of the MSTR are distinguished by two multisets of dimensions 
 \begin{equation}
\dim A=\langle\dim a\rangle_{a\in A},\quad \dim B=\langle\dim b\rangle_{b\in B}.
\end{equation}
 The MSTR are not symmetric with respect to the exchange of $A$ and $B$. For example, there are no known solutions for $\dim A=\langle2\rangle$ and $\dim B=\langle1^4\rangle$, while the two examples below give the simplest solutions for $\dim A=\langle1^4\rangle$ and $\dim B=\langle2\rangle$, one constant and the other rapidity-dependent.\hfill$\diamond$
\end{remark}
\begin{example}\label{ex:const-4}
 The simplest solution of the constant MSTR that is neither a solution of the STR nor a solution of the YBE corresponds to the dimensions  $\dim A=\langle1^4\rangle$ and $\dim B=\langle2\rangle$. We assume that $\operatorname{char}(\Bbbk)\ne 2$ and $\ri = \sqrt{-1} \in \Bbbk.$ 
 
 Let $K_4=(\mathbb{Z}/2\mathbb{Z})^2$ be the Klein four-group, for which we use the set-theoretic bijection 
 \begin{equation}
|\cdot|\colon K_4\to\{0,1,2,3\},\quad x\mapsto |x|=x_1+2x_2,
\end{equation}
where $\mathbb Z/2\mathbb Z$ is identified with the two-element set $\{0,1\}$. 
 We take $A=\{A_x\}_{x\in K_4}$ to be the set of one-dimensional vector spaces $A_x\simeq\Bbbk$, indexed by elements of $K_4$, and $B=\{B_0\}$ to be the singleton consisting of a two-dimensional vector space $B_0\simeq\Bbbk^2$.

 Consider the following bicharacter on $K_4$:
 \begin{equation}\label{eq:bichar}
\chi(x,y)=(-1)^{x_1y_1+x_2y_2+x_1y_2}.
\end{equation}
 A solution of the MSTR~\eqref{eq:mstr-alg} is given by 
\begin{equation}
 w_{A_x,A_y}=\chi(x,y-x),\quad x,y\in K_4,
\end{equation}
which can be arranged in the following scalar matrix:
 \begin{equation}
W=(w_{A_x,A_y})_{x,y\in K_4}=\begin{pmatrix}
 1&1&1&1\\-1&1&1&-1\\-1&-1&1&1\\-1&1&-1&1
 \end{pmatrix},
\end{equation}
where the rows and columns are ordered by $|x|$,
together with four invertible $2\times 2$ matrices 
 \begin{equation}
v_{B_0,A_x}=\sigma_0+\ri P_x\in \operatorname{M}_2(\Bbbk)\simeq\operatorname{End}(B_0),\quad x\in K_4,
\end{equation}
  where 
\begin{equation}
P_{0}=\sum_{j=1}^3\sigma_j,\quad P_x=2\sigma_{|x|}-P_{0},\quad x\ne 0,
\end{equation}
with the Pauli matrices
\begin{equation}\label{eq:Pauli-mat}
 \sigma_0=\begin{pmatrix}1&0\\0&1\end{pmatrix},\quad \sigma_1=\begin{pmatrix}0&1\\1&0\end{pmatrix},\quad
 \sigma_2=\begin{pmatrix}0&-\ri\\ \ri&0\end{pmatrix},\quad
 \sigma_3=\begin{pmatrix}1&0\\0&-1\end{pmatrix}
\end{equation}
forming a basis of $\operatorname{M}_2(\Bbbk)$.
\hfill$\diamond$
\end{example}
\begin{example}\label{ex:rapidity-4} This example gives a rapidity-dependent solution of the MSTR corresponding to the $\mathfrak{sl}_3$ chiral Potts model at the only root of unity of order two. The regions are labelled
by the same sets of vector spaces $A$ and $B$ as in Example~\ref{ex:const-4}, and we assume $\Bbbk=\mathbb{C}$. 

 Choose a five-element subset $S=\{0,1,\lambda,\mu,\nu\}\subset\Bbbk$ and fix roots
 \begin{equation}
\sqrt{1-\lambda},\quad \sqrt{\mu-\nu},\quad \sqrt{\mu-\lambda},\quad \sqrt{-\nu}.
\end{equation}
Define the rapidity curve 
\begin{equation}
\rc=\{(z,u_0,u_1,u_\lambda,u_\mu,u_\nu)\in\Bbbk^6\mid \alpha+u_\alpha^2=z,\ \alpha\in S\},
\end{equation}
and choose a point $o\in\rc$ such that $z(o)=0.$

Define elements $c_k,f_k\in\Bbbk(\rc^2)$, $1\le k\le3$, by
\begin{equation}\label{eq:abc}
c_1=\frac{\sqrt{1-\lambda}z^-}{u^+_{1,\lambda}u^+_0},\quad
c_2=\frac{\sqrt{1-\lambda}\sqrt{\mu-\nu}z^-}{u^+_{1,\lambda}u_{\mu,\nu}},\quad
c_3=\frac{\sqrt{\mu-\nu}z^-}{u^+_{\mu,\nu}u^+_0},
\end{equation}
\begin{equation}
 f_1=\frac{u^-_{\mu,\lambda}u^-_{0,\nu}}{\sqrt{\mu-\lambda}\sqrt{-\nu}z^-},\quad f_2=\frac{u^-_{1}u^-_{0,\nu}}{\sqrt{-\nu}z^-},\quad f_3=\frac{u^-_{\mu,\lambda}u^-_1}{\sqrt{\mu-\lambda}z^-},
\end{equation}
where, for $p,q\in\rc$ and $\alpha,\beta\in S$,
\begin{equation}
z^-(p,q)=z(p)-z(q),\quad u^\pm_{\alpha,\beta}(p,q)=u_\alpha(p)u_\beta(q)\pm u_\beta(p)u_\alpha(q),\quad u^\pm_\alpha(p,q)=u_\alpha(p)\pm u_\alpha(q).
\end{equation}
The solution for the $w$-weights is of the form
\begin{equation}
w_{A_x,A_y}=c_{|x-y|}\chi(x,y-x),\quad x,y\in K_4,
\end{equation}
where $c_0=1$ and the bicharacter $\chi(x,y)$ is defined in~\eqref{eq:bichar}. This can also be written in matrix form
\begin{equation}\label{eq:F}
W=(w_{A_x,A_y})_{x,y\in K_4}=\begin{pmatrix}
1&c_1&c_2&c_3\\-c_1&1&c_3&-c_2\\-c_2&-c_3&1&c_1\\-c_3&c_2&-c_1&1
\end{pmatrix}\in\operatorname{M}_4\big(\Bbbk(\rc^2)\big),
\end{equation}
where the rows and columns are ordered by $|x|$. The $v$-weights of the solution are of the form
\begin{equation}\label{eq:V}
v_{B_0,A_x}=V_x:=\gamma(\sigma_0+\ri P_x),\quad x\in K_4,
\end{equation}
where
\begin{equation}
P_{0}=\mathbf f\cdot \bm\sigma:=\sum_{j=1}^3f_j\sigma_j,\quad P_x=2f_{|x|}\sigma_{|x|}-P_{0},\quad x\ne 0,
\end{equation}
with the Pauli matrices defined in~\eqref{eq:Pauli-mat}, and
$\gamma\in \Bbbk(\rc^2)$ is defined by
\begin{equation}\label{eq:gamma-klein}
\gamma(p,q)=\frac14\Big(
1-\mathbf c(p,q)\cdot\mathbf c(q,o)+
\frac{\mathbf c(p,q)+\mathbf c(q,o)-\mathbf c(p,q)\times\mathbf c(q,o)}{\mathbf c(p,o)}
\Big),
\end{equation}
where $\mathbf c=(c_1,c_2,c_3)$ is a three-component vector, and we use the usual dot and cross products with the additional convention
$$
\frac{\mathbf x}{\mathbf y}:=\sum_{j=1}^3 \frac{x_j}{y_j}.
$$
The composition law
\begin{equation}\label{eq:composition}
V_x(q,r)V_y(p,q)=\sum_{m\in K_4}
\frac{w_{x,m}(p,q)w_{m,y}(q,r)}{w_{x,y}(p,r)}
V_m(p,r),\quad x,y\in K_4,
\end{equation}
is an equivalent form of the MSTR~\eqref{eq:MSTR-rapidity}.
\end{example}
\section{Solutions of the MSTR underlying the $\mathfrak{sl}_n$ chiral Potts models}
In this section, our goal is to construct a rapidity-dependent solution to the MSTR based on the $\mathfrak{sl}_n$ chiral Potts model. To this end, following~\cite{BKMS1991}, we start by recalling the definitions of local states and the rapidity curve of the $\mathfrak{sl}_n$ chiral Potts model. We assume that $\Bbbk=\mathbb{C}$.

Fix an integer $n\in\Z_{\ge2}$. Let $\omega\in\Bbbk$ be a root of unity of order $N=\operatorname{ord}(\omega)\ge2$. Set 
\begin{equation}
G=(\zm{N})^n/\langle(1,\ldots,1)\rangle\simeq (\zm{N})^{n-1},\quad M=N^{n-1}=|G|.
\end{equation}
We will think of elements of $(\zm{N})^n$ as maps $x\colon \zm{n}\to \zm{N}$, with the notation $x_i$ for the image of $i\in\zm{n}$. The following bicharacter on 
$(\zm{N})^n$ descends to a well-defined non-degenerate bicharacter on $G$:
\begin{equation}
 \chi(x,y)=\omega^{\sum_{i\in\zm{n}}(x_{i-1}-x_i)y_i}=\omega^{\sum_{i\in\zm{n}}x_i(y_{i+1}-y_i)},
\end{equation}
The two expressions show that $\chi$ is well defined in both arguments; its nondegeneracy follows directly from the formula. Note that $\chi$ is symmetric, that is, $\chi(x,y)=\chi(y,x)$, if and only if $n=2$, a property which is crucial for the Boltzmann weights of the chiral Potts model of Baxter--Perk--Au-Yang to satisfy the usual STR.

 Generalizing Examples~\ref{ex:const-4} and~\ref{ex:rapidity-4}, we define the set $A$ in Definition~\ref{def:mstr} as a set of $1$-dimensional vector spaces indexed by the group $G$:
 \begin{equation}
A=\{A_x\simeq\Bbbk\mid x\in G\}.
\end{equation}

\subsection{The rapidity curve and ratio functions}
Fix a family of \emph{moduli matrices} 
\begin{equation}
\{K_{i}\mid i\in\zm{n}\}\subset SL_2(\Bbbk),\quad K_0=I_2,
\end{equation}
and scalars $\kappa_i\in\Bbbk$, $i\in\zm{n}$, such that 
\begin{equation}
\kappa^N_i=(K_{i+1})_{1,1}(K_i)_{2,2}-(K_{i+1})_{1,2}(K_i)_{2,1}\ne0,\quad i\in\zm{n}.
\end{equation}
Define a projective \emph{rapidity curve} $\rc\subset \mathbb{P}^{2n-1}$ by
\begin{equation}
\rc=\big\{h^\pm\in \Bbbk^{n}\mid \big(\begin{smallmatrix}
 (h^+_i)^N\\
  (h^-_i)^N
\end{smallmatrix}\big)=K_{i}\big(\begin{smallmatrix}
 (h^+_0)^N\\
  (h^-_0)^N
\end{smallmatrix}\big),\ i\in\zm{n}
\big\},
\end{equation}
Here, for convenience, the coordinates $h^\pm_i\in\Bbbk$ are indexed by elements of $\zm{n}$. 

The \emph{ratio functions} $\xi_i,\zeta_i\in\Bbbk(\rc)$, defined by
\begin{equation}
\xi_i:=\frac{h^+_i}{h^-_0},\quad\zeta_i:=\frac{h^-_i}{h^-_0},\quad  i\in\zm{n},
\end{equation}
satisfy the relations
\begin{equation}\label{eq:rc-ratio}
\begin{pmatrix}
 \xi_i^N\\
 \zeta_i^N
\end{pmatrix}
=K_i\begin{pmatrix}
 \xi_0^N\\
1
\end{pmatrix},\quad  i\in\zm{n}.
\end{equation}
Since $\zeta_0=1$, there are $2n-1$ nontrivial ratio functions. For any $f\in\Bbbk(\rc)$, the two elements $\dot{f},\ddot{f}\in\Bbbk(\rc^2)$
are defined by the pullbacks
\begin{equation}
\dot{f}:=\operatorname{pr}_1^*\!f,\quad \ddot{f}:=\operatorname{pr}_2^*\!f,
\end{equation}
where $\operatorname{pr}_k\colon \rc^2\to\rc$ is the projection onto the $k$-th factor. Similarly, for the three projections
$\operatorname{pr}_k\colon \rc^3\to\rc$, $1\le k\le 3$, we keep the same notation for the first two pullbacks, and 
\begin{equation}
\dddot{f}:=\operatorname{pr}_3^*\!f.
\end{equation}

For $i\in\zm{n}$, define elements $\rho_i, \tau_i,\hat \rho_i, \hat\tau_i\in\Bbbk(\rc^2)$  by
\begin{equation}
 \rho_i=\frac{\dot{\zeta}_i\ddot{\xi}_i}{\dot{\xi}_i\ddot{\zeta}_i},\quad
 \tau_i=\frac{\dot{\xi}_i\ddot{\zeta}_i}{\dot{\xi}_0},\quad \hat\rho_i=\frac{\dot{\xi}_{i+1}\ddot{\zeta}_i}{\dot{\zeta}_i\ddot{\xi}_{i+1}},\quad\hat\tau_i=\frac{\dot{\zeta}_i\ddot{\xi}_{i+1}}{\kappa_i\ddot{\xi}_0}.
\label{eq:ratios}
\end{equation}
%
For any $i\in\zm{n}$, equations~\eqref{eq:rc-ratio} imply that
\begin{equation} \label{eq:curve}
 \tau_i^N=
 \frac{1-\rho_0^N}{1-\rho_i^N},
\quad 
 \hat\tau_i^N=
 \frac{1-\rho_0^{-N}}{1-\hat\rho_i^N}.
\end{equation}
Indeed, if $K_i=\big(
\begin{smallmatrix}
 a&b\\c&d
\end{smallmatrix}\big)
$ and $K_{i+1}=\big(
\begin{smallmatrix}
 a'&b'\\c'&d'
\end{smallmatrix}\big)$, then using~\eqref{eq:rc-ratio} and denoting $\dot{u}:=\dot{\xi}^N_0$ and $\ddot{u}:=\ddot{\xi}^N_0$, we compute
\begin{multline}
\frac{\tau_i^N-\frac{1-\rho_0^N}{1-\rho_i^N}}{(a+b\dot{u}^{-1})(c\ddot{u}+d)}
= 1
-\frac{\dot{u}-\ddot{u}}{(a\dot{u}+b)(c\ddot{u}+d)-(a\ddot{u}+b)(c\dot{u}+d)}\\
= 1
-\frac{\dot{u}-\ddot{u}}{(ad-bc)(\dot{u}-\ddot{u})}= 1
-\frac{1}{ad-bc}=0
\end{multline}
and
\begin{multline}
 \frac{\Big( \hat\tau_i^N-
 \frac{1-\rho_0^{-N}}{1-\hat\rho_i^N}\Big)(a'd-b'c)}{(c\dot{u}+d)(a'+b'\ddot{u}^{-1})}=1-\frac{(\ddot{u}-\dot{u})(a'd-b'c)}{(c\dot{u}+d)(a'\ddot{u}+b')-(c\ddot{u}+d)(a'\dot{u}+b')}\\
 =1-\frac{(\ddot{u}-\dot{u})(a'd-b'c)}{(da'-cb')(\ddot{u}-\dot{u})}=0.
\end{multline}

\subsection{Dual weights}
For $m\ge0$, the standard $\omega$-Pochhammer symbol $(z;\omega)_m$ is defined by
\begin{equation}\label{eq:poch}
 (z;\omega)_m=\prod_{j=0}^{m-1}(1-z\omega^j)=(z\omega^{m-1};\omega^{-1})_m.
 \end{equation}
 It extends to all $m\in \Z$ via the identity
 \begin{equation}
(z;\omega)_{m+k}=(z;\omega)_m(z\omega^m;\omega)_k.
\end{equation}
In particular, for $m=-k$, we obtain
\begin{equation}
1=(z;\omega)_{-k}(z\omega^{-k};\omega)_k
\end{equation}
or equivalently
\begin{equation}
\label{eq:inv-rels-poch-w}
1=(z;\omega)_{-k}(z\omega^{-1};\omega^{-1})_k.
\end{equation}

For any $x\in G$, represented by a function $x\colon\zm{n}\to\Z$, define \emph{dual weights} $c_x,\hat c_x\in\Bbbk(\rc^2)$ by
\begin{align}
 c_x&=\prod_{i\in\zm{n}}
\tau_i^{x_{i+1}-x_{i}}
 (\rho_i;\omega^{-1})_{x_{i+1}-x_{i}},\label{eq:a}\\
 \hat c_x&=\prod_{i\in\zm{n}}
\hat\tau^{x_{i+1}-x_i}_i
(
 \hat\rho_i;\omega)_{x_{i+1}-x_i}.
\label{eq:dual}
\end{align}
By~\eqref{eq:curve}, the local factor in each product is quasi-periodic, with constant multiplier:
\begin{equation}
 \tau_i^{k+N}(\rho_i;\omega^{-1})_{k+N}
 =(1-\rho_0^N)\tau_i^{k}(\rho_i;\omega^{-1})_{k},
\end{equation}
and similarly for the second product. When a single component of $x$ is increased by $N$, two adjacent differences are shifted by $N$ in opposite directions, so the multipliers cancel. Uniform shifts of all components leave the differences unchanged. Thus $c_x$ and $\hat c_x$ are well-defined elements of $\Bbbk(\rc^2)$ for each $x\in G$. Their restrictions to the diagonal satisfy
\begin{equation}
\Delta^*c_x=\Delta^*\hat c_x=\delta_{0,x},\quad \forall x\in G,
\end{equation}
where $\Delta\colon \rc\hookrightarrow\rc^2$ is the diagonal embedding.

The $w$-weights of the MSTR are elements of $\Bbbk(\rc^2)$ defined by
\begin{equation}w_{A_x,A_y}=W_{x,y}:=c_{y-x}\chi(x,y-x).\label{eq:w}
\end{equation}
The dual weights $\hat c_x$ will be used below to define the $v$-weights.
\begin{remark}
The relation between the weights in~\eqref{eq:w} and the two-spin Boltzmann weights $\overline{W}_{p,q}(k,l)$ of~\cite{BKMS1991} is given by the equality
\begin{equation}
W_{x,y}(p,q)=\overline{W}_{p,q}(y,x),
\end{equation}
where $x$ and $y$ are interchanged because of the different graphical assignments of spin variables relative to the rapidity lines in the MSTR and in~\cite{BKMS1991}.
\hfill$\diamond$
\end{remark}

\subsection{Fourier transformation of the MSTR}
For an element $f\in\Bbbk(\rc^2)$, we adopt the convention that the elements
$f',f'',f'''\in\Bbbk(\rc^3)$ are the pullbacks of $f$ corresponding to omitting the first, second, and third arguments, respectively.

With this convention, the MSTR~\eqref{eq:mstr} is equivalent to
\begin{equation}\label{eq:comp}
 v'_{b,A_x}v'''_{b,A_y}=\sum_{m\in G} C^m_{x,y}v''_{b,A_m},\quad
 C^m_{x,y}=\frac{W'''_{x,m}W'_{m,y}}{W''_{x,y}},
\end{equation}
Here the set $B$ of vector spaces, and hence its elements $b$, remains to be determined.

The elements $C^m_{x,y}\in\Bbbk(\rc^3)$ have the translation invariance property 
\begin{equation}
C^{m+z}_{x+z,y+z}=C^m_{x,y},
\end{equation}
which follows from the substitution of~\eqref{eq:w} into their definition:
\begin{equation}
\frac{C^m_{x,y}c''_{y-x}}{c'''_{m-x}c'_{y-m}}=\frac{\chi(x,m-x)\chi(m,y-m)}{\chi(x,y-x)}=\frac{\chi(m,y-m)}{\chi(x,y-m)}=\chi(m-x,y-m).
\end{equation}
Thus the Fourier components
\begin{equation}\label{eq:gbx}
 g_{b,x}=\frac1M\sum_{y\in G}\chi(x,y)^{-1}v_{b,A_y}
\end{equation}
satisfy
\begin{equation}\label{eq:gg=g}
 g'_{b,x}g'''_{b,y}=\mathcal{K}_{x,y}g''_{b,x+y},
\end{equation}
where $\mathcal{K}_{x,y}\in\Bbbk(\rc^3)$ are defined by
\begin{equation}\label{eq:K}
 \mathcal{K}_{x,y}=\frac1M\sum_{k,l\in G}
 \frac{C^0_{k,l}}{\chi(x,k)\chi(y,l)},\quad x,y\in G.
\end{equation}
Fix a generic point $o\in\rc$.
Define the element $\gamma\in\Bbbk(\rc^2)$ by
\begin{equation}\label{eq:gamma}
 \gamma(p,q)=\calK_{0,0}(p,q,o).
\end{equation}
This is an explicit rational function of the ratio functions by~\eqref{eq:a} and~\eqref{eq:w}; $\gamma(p,p)=1$ by continuation, and  $\gamma(p,o)=1$.
\begin{proposition}\label{prop:fourier-duality}
 The double sum in~\eqref{eq:K} can be evaluated with the result
\begin{equation}
 \calK_{x,y}=
 \frac{\gamma'''\gamma'}{\chi(y,x)\gamma''}\,
 \frac{\hat c'_x\hat c'''_y}{\hat c''_{x+y}}.\label{eq:factor}
\end{equation}
\end{proposition}
We first establish several preliminary identities needed for the proof.
\subsection{Cyclic analogue of Ramanujan's ${_1\psi_1}$-summation formula}
Following~\cite[Appendix A]{MR1210206}, we introduce a cyclic analogue of Ramanujan's ${_1\psi_1}$ basic hypergeometric function
\begin{equation}\label{eq:def-raman-cyc}
 \ram(a,b;\omega,z)=\sum_{j\in\zm{N}}
 \frac{(a;\omega)_j}{(\omega b;\omega)_j}z^j,
 \quad z^N=\frac{1-b^N}{1-a^N},
\end{equation}
where, compared to~\cite[(A.1)]{MR1210206}, we use slightly different conventions and notation. The function $f(x,y\mid z)$ in~\cite[(A.1)]{MR1210206}
is $\ram(\omega y,x;\omega,z)$ in our notation. In particular, with definition~\eqref{eq:def-raman-cyc}, the following symmetry holds:
\begin{equation}\label{eq:symmetry-psi-raman}
\ram(a,b;\omega,z)=\ram(b,a;\omega^{-1},z^{-1}),
\end{equation}
which is easily verified by using~\eqref{eq:inv-rels-poch-w}:
\begin{equation}
\ram(a,b;\omega,z)=\sum_{j\in\zm{N}}
 \frac{(a;\omega)_j}{(\omega b;\omega)_j}z^j=\sum_{j\in\zm{N}}
 \frac{(b;\omega^{-1})_{-j}}{(\omega^{-1} a;\omega^{-1})_{-j}}z^j=\ram(b,a;\omega^{-1},z^{-1}),
\end{equation}
where, in the last equality, we have changed the summation index $j\mapsto -j$.
\begin{lemma}[{cf.~\cite[(A.14)]{MR1210206}}]\label{lem:cyc-raman}
 For every
$k,l,m\in\Z$, we have
\begin{equation}\label{eq:H-master}
 \frac{\ram(\omega^k a,\omega^l b;\omega,\omega^m z)}{\ram(a,b;\omega, z)}
 =
 \frac{(\omega b;\omega)_l(\frac a b;\omega)_{k-l}(z;\omega)_m}
 {\omega^{lm}z^{l}b^{m}(a;\omega)_k(\frac{az}b;\omega)_{k-l+m}}.
\end{equation}
\end{lemma}
The proof of~\cite[Appendix A]{MR1210206}, adapted to our notation, is given in Appendix~\ref{app:cyc-raman}.
\begin{remark} Equation~\eqref{eq:H-master} can be written in many different ways. For example, by using the symmetry~\eqref{eq:symmetry-psi-raman}, we write
\begin{equation}\label{eq:H-master-dif-forms-1}
 \frac{\ram(\omega^l b,\omega^k a;\omega^{-1},\omega^{-m} z^{-1})}{\ram(b,a;\omega^{-1}, z^{-1})}
 =
 \frac{(\omega b;\omega)_l(\frac ab;\omega)_{k-l}(z;\omega)_m}
 {\omega^{lm}z^{l}b^{m}(a;\omega)_k(\frac{az}b;\omega)_{k-l+m}},
 \end{equation}
 where we can make the substitutions $k\leftrightarrow -l$, $a\leftrightarrow b$, $z\mapsto z^{-1}$, and $\omega\mapsto \omega^{-1}$:
 \begin{equation}\label{eq:H-master-dif-forms-2}
 \frac{\ram(\omega^k a,\omega^l b;\omega,\omega^m z)}{\ram(a,b;\omega, z)}
 =
 \frac{(\omega^{-1} a;\omega^{-1})_{-k}(\frac ba;\omega^{-1})_{k-l}(z^{-1};\omega^{-1})_m}
 {\omega^{km}z^{k}a^{m}(b;\omega^{-1})_{-l}(\frac b{az};\omega^{-1})_{k-l+m}},
 \end{equation}
 and by using~\eqref{eq:inv-rels-poch-w}
\begin{equation}\label{eq:H-master-dif-forms-3}
 \frac{\ram(\omega^k a,\omega^l b;\omega,\omega^m z)}{\ram(a,b;\omega, z)}
 =
 \frac{(\omega b;\omega)_l(\frac{\omega b}{az};\omega)_{l-k-m}(z^{-1};\omega^{-1})_m}
 {\omega^{km}z^{k}a^{m}(a;\omega)_{k}(\frac{\omega b}a;\omega)_{l-k}}.
 \end{equation}
 The special case $k=0$ of~\eqref{eq:H-master-dif-forms-3} will be particularly useful for us:
 \begin{equation}\label{eq:H-master-dif-forms-4}
 \frac{\ram(a,\omega^l b;\omega,\omega^m z)}{\ram(a,b;\omega, z)}
 =
 \frac{(\omega b;\omega)_l(\frac{\omega b}{az};\omega)_{l-m}(z^{-1};\omega^{-1})_m}
 {a^{m}(\frac{\omega b}a;\omega)_{l}}.
 \end{equation}
\hfill$\diamond$
\end{remark}
\subsection{Bracket function}
To a map $f\colon G\to \Bbbk(\rc^2)$, $x\mapsto f_x$, and a bicharacter $\psi$ on $G$, we associate
a map $[f,\psi]\colon G^2\to \Bbbk(\rc^3)$, $(x,y)\mapsto [f,\psi]_{x,y}$, defined by
\begin{equation}\label{eq:partial}
 [f,\psi]_{x,y}
 =f'_y\sum_{m\in G}\frac{f'''_m}{f''_{m+y}}
 \psi(x,m)\psi(m,y).
\end{equation}
Using this map, we write 
\begin{equation}\label{eq:1-sum-calcul}
\sum_{k\in G}\chi(x,k)^{-1}C^0_{k,y}=\sum_{k\in G}\frac{c'''_{-k}c'_{y}}{\chi(x,k)\chi(k,y)c''_{y-k}}=c'_{y}\sum_{k\in G}\frac{c'''_{k}}{c''_{k+y}}\chi(x,k)\chi(k,y)= [c,\chi]_{x,y}.
\end{equation}

\begin{lemma}
 Let the field $\Bbbk(\rc^3)$ be extended by adjoining an algebraic element $\alpha$ satisfying 
 \begin{equation}
\alpha^N=\frac{1-(\rho''_0)^N}{1-(\rho'''_0)^N}.
\end{equation}
Then the following identity holds:
 \begin{equation}\label{eq:original-H}
 [c,\chi]_{x,y}=\frac{\Phi_\omega}{N}\sum_{h\in\zm{N}}\prod_{i\in\zm{n}}\frac{\big(\frac{\rho'_i\tau''_i}{\alpha\tau'''_i};\omega^{-1}\big)_{x_i-y_{i}+h}\big(\frac{\tau''_i}{\alpha\tau'''_i};\omega\big)_{y_{i+1}-x_i-h}}
 {\big(\frac{\tau''_i}{\tau'_i}\big)^{x_i-y_{i}+h}\big(\frac{\tau''_i\rho'''_i}{\tau'_i}\big)^{y_{i+1}-x_i-h}},
\end{equation}
where
\begin{equation}\label{eq:BigPhi}
\Phi_\omega:=\prod_{i\in\zm{n}}\ram\Big(\rho'''_i,\rho''_i\omega;\omega^{-1},\frac{\tau'''_i}{\tau''_i}\alpha\Big).
\end{equation}
\end{lemma}
\begin{proof}
 Substituting~\eqref{eq:a} into $[c,\chi]_{x,y}$,
we write
\begin{equation}
 [c,\chi]_{x,y}=\frac{c'_y}{c''_y}\sum_{m\in G}\prod_{i\in\zm{n}}\frac{(\rho'''_i;\omega^{-1})_{m_{i+1}-m_{i}}}{(\rho''_i\omega^{y_{i}-y_{i+1}};\omega^{-1})_{m_{i+1}-m_{i}}}\Big(\frac{\tau'''_i}{\tau''_i}\alpha\omega^{x_i-y_{i+1}}\Big)^{m_{i+1}-m_{i}},
\end{equation}
where we have inserted the element $\alpha$ so that the local factor in the product is now $N$-periodic, rather than only quasi-periodic in each $m_i$. For each $m\in G$, we can choose its unique representative with $m_0=0$ and the summation runs over the other components. Due to the $N$-periodicity of each local factor,  we can introduce new summation variables  $d\colon\zm{n}\to \zm{N}$, $i\mapsto d_i$, satisfying the equations
\begin{equation}\label{eq:d-m-eqns}
d_i=m_{i+1}-m_{i},\quad i\in\zm{n},
\end{equation}
so that they are constrained by the condition
\begin{equation}
\sum_{i\in\zm{n}}d_i=0,
\end{equation}
which means that there are only $n-1$ independent components $d_0,d_1,\dots,d_{n-2}$. Solving~\eqref{eq:d-m-eqns} for the $m_i$ (with $m_0=0$), we obtain
\begin{equation}
m_i=\sum_{j=0}^{i-1}d_j,\quad 1\le i\le n-1,
\end{equation}
and thus
\begin{equation}
 [c,\chi]_{x,y}=\frac{c'_y}{c''_y}\sum_{d_0,d_1,\dots,d_{n-2}\in\zm{N}}\prod_{i\in\zm{n}}\frac{(\rho'''_i;\omega^{-1})_{d_i}}{(\rho''_i\omega^{y_{i}-y_{i+1}};\omega^{-1})_{d_i}}\Big(\frac{\tau'''_i}{\tau''_i}\alpha\omega^{x_i-y_{i+1}}\Big)^{d_i},
\end{equation}
where $d_{n-1}=-\sum_{i=0}^{n-2}d_i$. We can restore the symmetry over all components of $d$ by inserting the group delta-function $\delta_{\zm{N}}(\sum_{i\in\zm{n}}d_i)$ and summing over all components of $d$. Furthermore, we can make all components of $d$ independent at the cost of one additional summation, using the formula
\begin{equation}
\delta_{\zm{N}}\Big(\sum_{i\in\zm{n}}d_i\Big)=\frac{1}{N}\sum_{h\in\zm{N}}\omega^{h\sum_{i\in\zm{n}}d_i}.
\end{equation}
Thus we obtain
\begin{align}
 \nonumber [c,\chi]_{x,y}&=\frac{c'_y}{Nc''_y}\sum_{h\in\zm{N}}\sum_{d\in(\zm{N})^n}\prod_{i\in\zm{n}}\frac{(\rho'''_i;\omega^{-1})_{d_i}}{(\rho''_i\omega^{y_{i}-y_{i+1}};\omega^{-1})_{d_i}}\Big(\frac{\tau'''_i}{\tau''_i}\alpha\omega^{h+x_i-y_{i+1}}\Big)^{d_i}\\
\nonumber&=\frac{c'_{y}}{Nc''_y}\sum_{h\in\zm{N}}\prod_{i\in\zm{n}}\sum_{d_i\in \zm{N}}\frac{(\rho'''_i;\omega^{-1})_{d_i}}{(\rho''_i\omega^{y_{i}-y_{i+1}};\omega^{-1})_{d_i}}\Big(\frac{\tau'''_i}{\tau''_i}\alpha\omega^{h+x_i-y_{i+1}}\Big)^{d_i}\\
 \label{eq:psi-prod}&=\frac{c'_{y}}{Nc''_y}\sum_{h\in\zm{N}}\prod_{i\in\zm{n}}\ram\Big(\rho'''_i,\rho''_i\omega^{y_i-y_{i+1}+1};\omega^{-1},\frac{\tau'''_i}{\tau''_i}\alpha\omega^{h+x_i-y_{i+1}}\Big),
\end{align}
where, in the last step, we have used definition~\eqref{eq:def-raman-cyc}. 

In each $\Psi$-function in~\eqref{eq:psi-prod}, we can use~\eqref{eq:H-master-dif-forms-4} to extract the dependence on the summation variable and the spin variables:
\begin{multline}
[c,\chi]_{x,y}=\frac{c'_{y}}{Nc''_y}\sum_{h\in\zm{N}}\prod_{i\in\zm{n}}\ram\Big(\rho'''_i,\rho''_i\omega;\omega^{-1},\frac{\tau'''_i}{\tau''_i}\alpha\Big)\\
 \times\frac{(\tau''_i)^{y_{i+1}-y_{i}}\big(\rho''_i;\omega^{-1}\big)_{y_{i+1}-y_{i}}\big(\frac{\rho'_i\tau''_i}{\alpha\tau'''_i};\omega^{-1}\big)_{x_i-y_{i}+h}\big(\frac{\tau''_i}{\alpha\tau'''_i};\omega\big)_{y_{i+1}-x_i-h}}
 {\big(\frac{\tau''_i}{\tau'_i}\big)^{y_{i+1}-y_{i}}(\rho'''_i)^{y_{i+1}-x_i-h}(\tau'_i)^{y_{i+1}-y_{i}}\big(\rho'_i;\omega^{-1}\big)_{y_{i+1}-y_{i}}},
\end{multline}
where we have used the following relations between the three different pullbacks:
\begin{equation}
\rho'_i\rho'''_i=\rho''_i,\quad i\in\zm{n},
\end{equation}
We have also inserted cancelling $\tau_i$-factors to identify the products corresponding to the two $c_y$-functions. Simplifying, using~\eqref{eq:BigPhi} and rearranging the remaining monomials in the denominator, we obtain~\eqref{eq:original-H}.
\end{proof}

\begin{lemma}\label{lem:bracket=bracket}
The following identity holds:
 \begin{equation}
\frac{ [c,\chi]_{x,y}}{\Phi_\omega}
 =\frac{[\hat c,\chi^{-1}]_{\sigma y,x}}{\widehat\Phi_{\omega^{-1}}},
\label{eq:target}
\end{equation}
where $\sigma\colon G\to G$ is a group automorphism defined by
\begin{equation}
(\sigma l)_i=l_{i+1},\quad i\in\zm{n},
\end{equation}
and
\begin{equation}\label{eq:BigHatPhi}
\widehat\Phi_{\omega^{-1}}:=\prod_{i\in\zm{n}}\ram\Big(\hat\rho'''_i,\hat\rho''_i\omega^{-1};\omega,\frac{\hat\tau'''_i}{\hat\tau''_i}\hat\alpha\Big).
\end{equation}
\end{lemma}
\begin{proof}
 The calculation of $[\hat c,\chi^{-1}]_{\sigma y,x}$ is obtained from that of $[c,\chi]_{x,y}$ by the substitutions
 \begin{equation}
(\rho,\tau, \alpha, x,y,\omega)\mapsto(\hat\rho,\hat\tau,\hat\alpha,\sigma y,x,\omega^{-1}),
\end{equation}
where 
\begin{equation}\label{eq:kms-alpha-hat}
\hat\alpha=\frac{\rho'''_0}{\rho''_0}\alpha=\frac{\ddot{\xi}_0}{\dddot{\xi_0}}\alpha,
\end{equation}
which satisfies the equation
\begin{equation}
\hat\alpha^N=\frac{1-(\rho''_0)^{-N}}{1-(\rho'''_0)^{-N}}.
\end{equation}
Thus, after these substitutions into~\eqref{eq:original-H}, we obtain
 \begin{equation}
\frac{N [\hat c,\chi^{-1}]_{\sigma y,x}}{\widehat\Phi_{\omega^{-1}}}=\sum_{h\in\zm{N}}\prod_{i\in\zm{n}}\frac{\big(\frac{\hat\tau''_{i-1}}{\hat\alpha\hat\tau'''_{i-1}};\omega^{-1}\big)_{x_{i}-y_{i}+h}\big(\frac{\hat\rho'_i\hat\tau''_i}{\hat\alpha\hat\tau'''_i};\omega\big)_{y_{i+1}-x_{i}-h}}
 {\big(\frac{\hat\tau''_{i-1}\hat\rho'''_{i-1}}{\hat\tau'_{i-1}}t\big)^{x_{i}-y_{i}+h}\big(\frac{\hat\tau''_i}{\hat\tau'_i}t\big)^{y_{i+1}-x_{i}-h}},
\end{equation}
Here we have replaced $h$ by $-h$, reordered the factors in the numerator and denominator, and shifted the product index in one factor of each. We have also inserted an arbitrary nonzero parameter $t$ in the denominator; its contributions cancel in the product. We now verify that, for a suitable choice of $t$, the resulting expression coincides with that in~\eqref{eq:original-H}:
\begin{equation}
\frac{\hat\tau''_{i-1}}{\hat\alpha\hat\tau'''_{i-1}}=\frac{\dddot{\xi_0}}{\alpha\ddot{\xi}_0}\frac{\dot{\zeta}_{i-1}\dddot{\xi_{i}}}{\kappa_{i-1}\dddot{\xi}_0}\frac{\kappa_{i-1}\ddot{\xi}_0}{\dot{\zeta}_{i-1}\ddot{\xi_{i}}}=\frac{\dddot{\xi_{i}}}{\alpha\ddot{\xi_{i}}},\quad
\frac{\rho'_i\tau''_i}{\alpha\tau'''_i}=\frac{\ddot{\zeta}_i\dddot{\xi_i}}{\alpha\ddot{\xi}_i\dddot{\zeta_i}}\frac{\dot{\xi}_i\dddot{\zeta_i}}{\dot{\xi}_0}\frac{\dot{\xi}_0}{\dot{\xi}_i\ddot{\zeta}_i}=\frac{\dddot{\xi_i}}{\alpha\ddot{\xi}_i},
\end{equation}
\begin{equation}
\frac{\hat\rho'_i\hat\tau''_i}{\hat\alpha\hat\tau'''_i}=\frac{\dddot{\xi_0}}{\alpha\ddot{\xi}_0}\frac{\ddot{\xi}_{i+1}\dddot{\zeta_i}}{\ddot{\zeta}_i\dddot{\xi}_{\!\!i+1}}\frac{\dot{\zeta}_i\dddot{\xi}_{\!\!i+1}}{\kappa_i\dddot{\xi_0}}\frac{\kappa_i\ddot{\xi}_0}{\dot{\zeta}_i\ddot{\xi}_{i+1}}=\frac{\dddot{\zeta_i}}{\alpha\ddot{\zeta}_i},\quad
\frac{\tau''_i}{\alpha\tau'''_i}=\frac{\dot{\xi}_i\dddot{\zeta_i}}{\alpha\dot{\xi}_0}\frac{\dot{\xi}_0}{\dot{\xi}_i\ddot{\zeta}_i}=\frac{\dddot{\zeta_i}}{\alpha\ddot{\zeta}_i},
\end{equation}
\begin{equation}
\frac{\hat\tau''_{i-1}\hat\rho'''_{i-1}}{\hat\tau'_{i-1}}t=\frac{\dot{\zeta}_{i-1}\dddot{\xi_{i}}}{\kappa_{i-1}\dddot{\xi_0}}\frac{\kappa_{i-1}\dddot{\xi_0}}{\ddot{\zeta}_{i-1}\dddot{\xi_{i}}}\frac{\dot{\xi}_{i}\ddot{\zeta}_{i-1}}{\dot{\zeta}_{i-1}\ddot{\xi}_{i}}t=\frac{\dot{\xi}_{i}}{\ddot{\xi}_{i}}t,\quad \frac{\tau''_i}{\tau'_i}=\frac{\dot{\xi}_i\dddot{\zeta_i}}{\dot{\xi}_0}\frac{\ddot{\xi}_0}{\ddot{\xi}_i\dddot{\zeta_i}}=\frac{\dot{\xi}_{i}\ddot{\xi}_0}{\ddot{\xi}_{i}\dot{\xi}_0},
\end{equation}
\begin{equation}
\frac{\hat\tau''_i}{\hat\tau'_i}t=\frac{\dot{\zeta}_i\dddot{\xi}_{\!\!i+1}}{\kappa_i\dddot{\xi_0}}\frac{\kappa_i\dddot{\xi_0}}{\ddot{\zeta}_i\dddot{\xi}_{\!\!i+1}}t=\frac{\dot{\zeta}_{i}}{\ddot{\zeta}_{i}}t,\quad \frac{\tau''_i\rho'''_i}{\tau'_i}=\frac{\dot{\xi}_{i}\ddot{\xi}_0}{\ddot{\xi}_{i}\dot{\xi}_0}\frac{\dot{\zeta}_i\ddot{\xi}_i}{\dot{\xi}_i\ddot{\zeta}_i}=\frac{\dot{\zeta}_{i}\ddot{\xi}_0}{\ddot{\zeta}_{i}\dot{\xi}_0}.
\end{equation}
We see that choosing $t=\ddot{\xi}_0/\dot{\xi}_0$ makes all factors agree, completing the proof.
\end{proof}
\subsection{Proof of Proposition~\ref{prop:fourier-duality}}
\begin{proof}
 By using definition~\eqref{eq:K} and the calculation in~\eqref{eq:1-sum-calcul}, we write
 \begin{equation}
 \mathcal{K}_{x,y}=\frac1M\sum_{k,l\in G}
 \frac{C^0_{k,l}}{\chi(x,k)\chi(y,l)}=\frac1M\sum_{l\in G}\frac{[c,\chi]_{x,l}}{\chi(y,l)}
\end{equation}
Using Lemma~\ref{lem:bracket=bracket} and definition~\eqref{eq:partial}, we compute
\begin{multline}
 \frac{\widehat\Phi_{\omega^{-1}}}{\Phi_\omega}\mathcal{K}_{x,y}=\sum_{l\in G}\frac{[\hat c,\chi^{-1}]_{\sigma l,x}}{M\chi(y,l)}=\sum_{l\in G}\frac{[\hat c,\chi^{-1}]_{\sigma l,x}\chi(\sigma l,y)}{M}=\sum_{l\in G}\frac{[\hat c,\chi^{-1}]_{l,x}\chi(l,y)}{M}\\
 =\sum_{l\in G}\frac{\chi(l,y)}{M}\hat c'_x\sum_{m\in G}\frac{\hat c'''_m}{\hat c''_{m+x} \chi(l,m)\chi(m,x)} =\frac{\hat c'_x\hat c'''_y}{\chi(y,x)\hat c''_{x+y}}.
\end{multline}
Putting $x=y=0$, we also obtain
\begin{equation}\label{eq:k00Phi-Phi}
\frac{\Phi_\omega}{\widehat\Phi_{\omega^{-1}}}=\mathcal{K}_{0,0}.
\end{equation}
Thus
\begin{equation}\label{eq:almost-final}
\mathcal{K}_{x,y}=\frac{\mathcal{K}_{0,0}\hat c'_x\hat c'''_y}{\chi(y,x)\hat c''_{x+y}}.
\end{equation}
It remains to determine $\mathcal K_{0,0}$. We do this directly from the
$N$-th power formula of~\cite[Appendix C, (C.7)]{MR1210206}.
Put
\begin{equation}\label{eq:kms-d}
 d_\omega(z):=\prod_{j=1}^{N-1}(1-z\omega^j)^j,
 \qquad N_2:=\frac{N(N-1)}2,\qquad \kappa:=\prod_{i\in\zm n}\kappa_i.
\end{equation}
In the conventions of~\eqref{eq:def-raman-cyc}, the KMS formula reads
\begin{equation}\label{eq:kms-psi-power}
 \ram(a,b;\omega,z)^N
 =a^{N_2}\frac{d_\omega(1)d_\omega(a/b)d_\omega(\omega b/(az))}
 {d_\omega(1/b)d_\omega(a)d_\omega(\omega/z)}.
\end{equation}
Indeed, the function $f(x,y\mid z)$ in~\cite[(A.1)]{MR1210206}
is $\ram(\omega y,x;\omega,z)$ in our notation.
We shall also use the elementary product identities
\begin{equation}\label{eq:kms-product-identities}
\begin{aligned}
 d_{\omega^{-1}}(z)&=(-z)^{N_2}d_\omega(z^{-1})\omega^{-\sum_{j=1}^{N-1}j^2},\\
 d_\omega(z)d_\omega(\omega/z)&=(-z)^{-N_2}(1-z^N)^{N-1}\omega^{\sum_{j=1}^{N-1}j(j+1)}.
\end{aligned}
\end{equation}
These follow by reversing the product indices. Define a rational function on $\rc^2$ by
\begin{equation}\label{eq:kms-Delta}
 F:=\kappa^{2N_2}\prod_{i\in\zm n}\frac{d_\omega(\omega\rho_i)}
 {d_\omega(\omega/\hat\rho_i)}\in\Bbbk(\rc^2).
\end{equation}
Applying~\eqref{eq:kms-psi-power} to the two products in~\eqref{eq:k00Phi-Phi}
 gives
\begin{equation}\label{eq:kms-power-expanded}
 \mathcal K_{0,0}^{N}
 =\prod_i\left[
 (\rho_i'''\hat\rho_i''')^{-N_2}
 \frac{d_\omega(\hat\rho_i''')}{d_\omega(1/\rho_i''')}
 \right]\frac{F'}{F''}.
\end{equation}
For completeness, the cancellation of the auxiliary algebraic element
$\alpha$ can be seen explicitly. Put
\[
 Z_i=\frac{\tau_i''}{\alpha\tau_i'''},\qquad
 Q_i=\rho_i'Z_i.
\]
Using~\eqref{eq:kms-product-identities}, the $N$-th powers of the local
factors of $\Phi_\omega$ and $\widehat\Phi_{\omega^{-1}}$ become, respectively,
\[
 (\omega\rho_i'')^{N_2}
 \frac{d_\omega(1)d_\omega(\omega\rho_i')d_\omega(1/Q_i)}
 {d_\omega(\omega\rho_i'')d_\omega(1/\rho_i''')d_\omega(\omega/Z_i)},
 \qquad
 (\hat\rho_i''')^{N_2}
 \frac{d_\omega(1)d_\omega(\omega/\hat\rho_i')d_\omega(Z_i)}
 {d_\omega(\omega/\hat\rho_i'')d_\omega(\hat\rho_i''')d_\omega(\omega Q_{i+1})}.
\]
Here we used
$\hat\rho_i''Q_{i+1}/(\omega\hat\rho_i''')=Z_i$ and
$\hat\alpha\hat\tau_i'''/\hat\tau_i''=1/Q_{i+1}$,
which follow from the ratio functions and~\eqref{eq:kms-alpha-hat}.
After cyclic reindexing, the remaining mixed factors are
\[
 \frac{d_\omega(\omega Q_i)d_\omega(1/Q_i)}
 {d_\omega(Z_i)d_\omega(\omega/Z_i)}
 =\Big(\frac{Z_i}{\omega Q_i}\Big)^{N_2}
 \Big(\frac{1-Q_i^N}{1-Z_i^N}\Big)^{N-1}
 =(\omega\rho_i')^{-N_2}(\rho_i''')^{-2N_2}.
\]
The last equality uses
$Z_i^N=(1-(\rho_i''')^N)/(1-(\rho_i'')^N)$ and
$\rho_i'\rho_i'''=\rho_i''$.
Combining the monomial factors gives~\eqref{eq:kms-power-expanded}.

The ratio functions also satisfy
\begin{equation}\label{eq:kms-ratio-products}
 \prod_i\rho_i\hat\rho_i=1,\qquad
 \prod_i\frac{1-\hat\rho_i^N}{1-\rho_i^{-N}}=\kappa^N.
\end{equation}
For the second identity, use
\[
 \frac{1-\hat\rho_i^N}{1-\rho_i^{-N}}
 =\Big(\frac{\rho_i\tau_i}{\rho_0\hat\tau_i}\Big)^N,
 \qquad
 \prod_i\frac{\rho_i\tau_i}{\rho_0\hat\tau_i}=\kappa;
\]
the latter product telescopes after substitution of the ratio functions.
Thus~\eqref{eq:kms-product-identities} gives
\begin{equation}\label{eq:kms-pair-factor}
 \prod_i(\rho_i\hat\rho_i)^{-N_2}
 \frac{d_\omega(\hat\rho_i)}{d_\omega(1/\rho_i)}
 =\kappa^{-2N_2}F\Big(\prod_i\rho_i\hat\rho_i\Big)^{-2N_2}
 \Big(\prod_i\frac{1-\hat\rho_i^N}{1-\rho_i^{-N}}\Big)^{N-1}
 =F.
\end{equation}
Substituting this into~\eqref{eq:kms-power-expanded}, we obtain
\begin{equation}\label{eq:K00power}
 \mathcal K_{0,0}^{N}=\frac{F'F'''}{F''}.
\end{equation}
This is an identity in $\Bbbk(\rc^3)$: the auxiliary roots have disappeared.
In particular, for $r=o$, definition~\eqref{eq:gamma} gives
\begin{equation}
 \gamma(p,q)^N=\frac{F(p,q)F(q,o)}{F(p,o)}.
\end{equation}
It follows that
\begin{equation}
 \Big(\mathcal K_{0,0}(p,q,r)
 \frac{\gamma(p,r)}{\gamma(q,r)\gamma(p,q)}\Big)^N=1.
\end{equation}
Consequently, the rational function
\[
 E:=\mathcal K_{0,0}\frac{\gamma''}{\gamma'\gamma'''}
 \in\Bbbk(\rc^3)
\]
satisfies $E^N=1$. Over $\Bbbk=\mathbb C$, in the function field of
an irreducible $\rc^3$, it is therefore a constant root of unity.
The diagonal determines this constant: by~\eqref{eq:K} and
$W_{k,0}(p,p)=\delta_{k,0}$,
\begin{equation}
 \mathcal K_{0,0}(p,p,r)
 =\frac1M\sum_{k,l\in G}
 \frac{\delta_{k,0}W_{0,l}(p,r)}{W_{k,l}(p,r)}
 =\frac1M\sum_{l\in G}1=1.
\end{equation}
Hence $\gamma(p,p)=1$ and
\begin{equation}
 E(p,p,r)=\frac{1\cdot\gamma(p,r)}{1\cdot\gamma(p,r)}=1.
\end{equation}
We conclude that $E=1$, or equivalently
\begin{equation}\label{eq:kms-scalar-final}
 \mathcal K_{0,0}=\frac{\gamma'''\gamma'}{\gamma''}.
\end{equation}
Together with~\eqref{eq:almost-final}, this proves~\eqref{eq:factor}.
\end{proof}

\subsection{Construction from Proposition~\ref{prop:fourier-duality}}\label{sec:blocks}
We construct the $v$-weights in six steps, working throughout in the
corresponding function fields of the rapidity curve.

\paragraph{\it Step 1: scalar weights and Fourier components.}
Keep $A=\{A_y\simeq\Bbbk\mid y\in G\}$ and
$W_{x,y}=c_{y-x}\chi(x,y-x)$ as in~\eqref{eq:w}. The Fourier transforms $g_{b,x}$ of the unknown matrix
weights $v_{b,A_y}$, defined in~\eqref{eq:gbx}, satisfy the relations~\eqref{eq:gg=g}, which are equivalent to the MSTR.

\paragraph{\it Step 2: normalization of the multiplication.}
Define
\begin{equation}\label{eq:construction-a}
a_x:=\gamma\hat c_x\chi(x,x).
\end{equation}
The bicharacter identity
\[
\chi(x+y,x+y)=\chi(x,x)\chi(y,y)\chi(x,y)\chi(y,x)
\]
and Proposition~\ref{prop:fourier-duality} give
\begin{equation}\label{eq:construction-K}
\mathcal K_{x,y}=\frac{a'_xa'''_y}{a''_{x+y}}\chi(x,y).
\end{equation}
Consequently, the ansatz $g_{b,x}=a_xR_{b,x}$ with constant matrices $R_{b,x}$
reduces the MSTR to
\begin{equation}\label{eq:construction-R}
R_{b,x}R_{b,y}=\chi(x,y)R_{b,x+y}.
\end{equation}

\paragraph{\it Step 3:  the fixed algebra.}
Define the twisted group algebra
\begin{equation}\label{eq:construction-algebra}
\mathcal A=\bigoplus_{x\in G}\Bbbk r_x,
\qquad r_xr_y=\chi(x,y)r_{x+y}.
\end{equation}
Its unit is $r_0$. Associativity follows from
\[
\chi(x,y)\chi(x+y,z)=\chi(y,z)\chi(x,y+z).
\]
Every representation $\pi_b\colon\mathcal A\to\operatorname{End}(b)$
therefore supplies matrices $R_{b,x}=\pi_b(r_x)$ satisfying
\eqref{eq:construction-R}.

Choose representatives with $x_0=0$ and let $e_1,\ldots,e_{n-1}$ be the
coordinate vectors. Choose $\sqrt\omega$ and put $\eta=(\sqrt\omega)^{N-1}$.
The generators $U_i:=\eta r_{e_i}$ satisfy
\begin{equation}\label{eq:alg}
U_i^N=1,\qquad U_iU_{i+1}=\omega U_{i+1}U_i,\qquad
U_iU_j=U_jU_i\quad(|i-j|>1).
\end{equation}
Indeed, $r_{e_i}^N=\omega^{-N(N-1)/2}=\eta^{-N}$ and
$\chi(e_i,e_{i+1})/\chi(e_{i+1},e_i)=\omega$.
The ordered monomials in the $U_i$ form a basis.

\paragraph{\it Step 4: irreducible blocks and the set $B$.}
The commutator bicharacter $\beta(x,y)=\chi(x,y)/\chi(y,x)$ has a radical
consisting of classes with $x_{i-1}=x_{i+1}$ for all $i\in\zm{n}$.
For odd $n$ these classes are constant, hence zero in $G$; for even $n$
they have alternating coordinates and form a cyclic group of order $N$.
Since $\Bbbk=\mathbb C$, the twisted group algebra is semisimple.
For odd $n$ it has one simple block; for even $n$ the central element
$U_1U_3\cdots U_{n-1}$ has the $N$ distinct eigenvalues $\omega^\nu$.
The explicit realizations below give
\begin{equation}\label{eq:construction-blocks}
\mathcal A\simeq
\begin{cases}
\operatorname{M}_{N^{(n-1)/2}}(\Bbbk),&n\text{ odd},\\
\displaystyle\bigoplus_{\nu\in\zm{N}}\operatorname{M}_{N^{(n-2)/2}}(\Bbbk),&n\text{ even}.
\end{cases}
\end{equation}
Take one irreducible module for each block. Thus
\begin{equation}
B=\{B_0\simeq\Bbbk^{N^{(n-1)/2}}\}
\end{equation}
 for odd $n$, and
\begin{equation}
B=\{B_\nu\simeq\Bbbk^{N^{(n-2)/2}}\mid\nu\in\zm{N}\}
\end{equation}
for even $n$.
In both cases $\sum_{b\in B}(\dim b)^2=M$.

\paragraph{\it Step 5: explicit constant matrices.}

Put $s=\lfloor (n-1)/2\rfloor$. On $(\Bbbk^N)^{\otimes s}$ define
\[
 X_j|a\rangle=\omega^{a_j}|a\rangle,\qquad Y_j|a\rangle=|a+e_j\rangle,
 \qquad X_0=I.
\]
Let $\nu=0$ for odd $n$, and $\nu\in \zm{N}$ for even $n$. An explicit $\nu$-th block realization is
\[
 U_{\nu,2j-1}=X_jX_{j-1}^{-1},\quad U_{\nu,2j}=Y_j\quad(1\leq j\leq s),
 \qquad U_{\nu,n-1}=\omega^\nu X_s^{-1}\quad(n=2s+2).
\]
For $n=2$ this means $N$ one-dimensional blocks $U_{\nu,1}=\omega^\nu$.

For $0\leq x_i<N$, define
\begin{align}
Q_{\nu,x}&=U_{\nu,n-1}^{x_{n-1}}\cdots U_{\nu,1}^{x_1},\\
R_{B_\nu,x}&=\eta^{-\sum_{i=1}^{n-1}x_i}
\omega^{\sum_{i=1}^{n-1}\binom{x_i}{2}}Q_{\nu,x}.
\label{eq:construction-explicit-R}
\end{align}
To check the scalar factor, note that
$r_{e_i}^{x_i}=\omega^{-\binom{x_i}{2}}r_{x_ie_i}$ and that multiplication
in decreasing index order introduces no further factors. Thus
$R_{B_\nu,x}=\pi_{B_\nu}(r_x)$, proving~\eqref{eq:construction-R}, including
when addition of coordinates requires reduction modulo $N$.
These matrices generate each full matrix block, since
$X_j=U_{\nu,1}U_{\nu,3}\cdots U_{\nu,2j-1}$ and $Y_j=U_{\nu,2j}$.
Together with the distinct central characters in the even case and the
dimension count, this verifies~\eqref{eq:construction-blocks}.

\paragraph{\it Step 6: Fourier inversion and the MSTR identity.}
Set
\begin{equation}\label{eq:vartheta}
\vartheta_x=\gamma\hat c_x\chi(x,x)
\eta^{-\sum_{i=1}^{n-1}x_i}
\omega^{\sum_{i=1}^{n-1}\binom{x_i}{2}}.
\end{equation}
Define the matrix $v$-weights by
\begin{equation}\label{eq:v}
v_{B_\nu,A_y}
=V_{\nu,y}:=\sum_{x\in G}\chi(x,y)a_xR_{B_\nu,x}
=\sum_{x\in G}\chi(x,y)\vartheta_xQ_{\nu,x}.
\end{equation}
There is no factor $1/M$ in this inverse transform. Its Fourier components
are $g_{B_\nu,x}=a_xR_{B_\nu,x}$, and
\begin{equation}
g'_{B_\nu,x}g'''_{B_\nu,y}
=a'_xa'''_yR_{B_\nu,x}R_{B_\nu,y}
=a'_xa'''_y\chi(x,y)R_{B_\nu,x+y}
=\mathcal K_{x,y}g''_{B_\nu,x+y}.
\end{equation}
Fourier inversion gives~\eqref{eq:comp}, or explicitly
\begin{equation}\label{eq:construction-mstr}
\sum_{m\in G}W'''_{x,m}V''_{\nu,m}W'_{m,y}
=V'_{\nu,x}W''_{x,y}V'''_{\nu,y},\quad x,y\in G,\quad \nu\in \frac{1+(-1)^n}{2}\,(\mathbb Z/N\mathbb Z).
\end{equation}
Thus the constructed weights satisfy the rapidity-dependent MSTR identity
\eqref{eq:MSTR-rapidity}.

\begin{remark}
The construction proves the MSTR identity. To obtain a solution in the full
sense of Definition~\ref{def:mstr}, one must additionally verify the
invertibility and dual-basis conditions required there. These are separate
non-degeneracy assertions, not consequences of the MSTR identity alone.
\hfill$\diamond$
\end{remark}

\section*{Acknowledgements}

I would like to thank Vladimir Mangazeev and Sergey Sergeev for discussions.  This work was supported in part by the granting of access to OpenAI’s models as part of the ChatGPT for Academic Researchers program. I acknowledge the use of Mathematica for assistance with symbolic calculations, and of ChatGPT (Open AI) for assistance with computations, derivations, and text editing.

\noindent{\bf Funding.} This work is partially supported by the SNSF research program NCCR The Mathematics of Physics (SwissMAP), the SNSF grants no.~200021-232258 and no.~200020-200400. 

\def\cprime{$'$} \def\cprime{$'$}
\providecommand{\bysame}{\leavevmode\hbox to3em{\hrulefill}\thinspace}
\providecommand{\MR}{\relax\ifhmode\unskip\space\fi MR }
\providecommand{\MRhref}[2]{%
  \href{http://www.ams.org/mathscinet-getitem?mr=#1}{#2}
}
\providecommand{\href}[2]{#2}

\appendix 
\section{Proof of Lemma~\ref{lem:cyc-raman}}\label{app:cyc-raman}
\begin{proof}
Here we adapt the proof of~\cite[Appendix A]{MR1210206} to our notation.
 The summand $h(j)=(a;\omega)_jz^j/(\omega b;\omega)_j$ is $N$-periodic, and
$h(j+1)(1-b\omega^{j+1})=zh(j)(1-a\omega^j)$.
Summing this recurrence gives
\begin{equation}\label{eq:psi-shift-z}
 \frac{\ram(a,b;\omega,\omega z)}{\ram(a,b;\omega, z)}
 =\frac{1-z}{b-az}.
\end{equation}
Replacing $z$ by $\omega^m z$ in~\eqref{eq:psi-shift-z}, we write
\begin{align}
\nonumber\frac{\ram(a,b;\omega,\omega^{m+1} z)}{\ram(a,b;\omega, \omega^{m}z)}
 &=\frac{1-\omega^{m}z}{b(1-\omega^{m}az/b)} =\frac{b^m(az/b;\omega)_{m}(z;\omega)_{m+1}}{b^{m+1}(az/b;\omega)_{m+1}(z;\omega)_{m}} \Leftrightarrow \\
\nonumber\frac{\ram(a,b;\omega,\omega^{m+1} z)b^{m+1}(az/b;\omega)_{m+1}}{(z;\omega)_{m+1}}&=\frac{\ram(a,b;\omega, \omega^{m}z)b^m(az/b;\omega)_{m}}{(z;\omega)_{m}}=\ram(a,b;\omega, z) \Leftrightarrow\\
\label{eq:zwm-iterated} \frac{\ram(a,b;\omega, \omega^{m}z)}{\ram(a,b;\omega, z)}&=\frac{(z;\omega)_{m}}{b^m(az/b;\omega)_{m}}.
\end{align}
Next, we shift $a$:
\begin{equation}
\ram(\omega a,b;\omega,z)=\sum_{j=0}^{N-1}
 \frac{(a;\omega)_j}{(\omega b;\omega)_j}\frac{1-\omega^ja}{1-a}z^j=\frac{\ram(a,b;\omega, z)-a\ram(a,b;\omega, \omega z)}{1-a},
\end{equation}
which, by using~\eqref{eq:psi-shift-z}, gives
\begin{equation}\label{eq:psi-shift-a}
 \frac{\ram(\omega a,b;\omega, z)}{\ram(a,b;\omega, z)}
 =\frac{b-a}{(1-a)(b-az)}.
\end{equation}
Replacing $a$ by $\omega^k a$ in~\eqref{eq:psi-shift-a}, we write
\begin{align}\nonumber
 \frac{\ram(\omega^{k+1} a,b;\omega, z)}{\ram(\omega^ka,b;\omega, z)}
 =\frac{1-\omega^ka/b}{(1-\omega^ka)(1-\omega^kaz/b)}& =\frac{(a/b;\omega)_{k+1}(a;\omega)_{k}(az/b;\omega)_{k}}{(a;\omega)_{k+1}(az/b;\omega)_{k+1}(a/b;\omega)_{k}}\Leftrightarrow  \\
 \nonumber \frac{\ram(\omega^{k+1} a,b;\omega, z)(a;\omega)_{k+1}(az/b;\omega)_{k+1}}{(a/b;\omega)_{k+1}}&=\frac{\ram(\omega^ka,b;\omega, z)(a;\omega)_{k}(az/b;\omega)_{k}}{(a/b;\omega)_{k}}=\ram(a,b;\omega, z)\Leftrightarrow\\
 \label{eq:awk-iterated} \frac{\ram(\omega^ka,b;\omega, z)}{\ram(a,b;\omega, z)}&=\frac{(a/b;\omega)_{k}}{(a;\omega)_{k}(az/b;\omega)_{k}}.
\end{align}
Replacing $z$ by $\omega^m z$ in~\eqref{eq:awk-iterated} and using~\eqref{eq:zwm-iterated}, we obtain
\begin{align}
\nonumber\frac{\ram(\omega^ka,b;\omega, \omega^m z)}{\ram(a,b;\omega,\omega^m z)}&=\frac{(a/b;\omega)_{k}}{(a;\omega)_{k}(\omega^maz/b;\omega)_{k}}\Leftrightarrow \\
\nonumber\frac{\ram(\omega^ka,b;\omega, \omega^m z)}{\ram(a,b;\omega,z)}&=\frac{(a/b;\omega)_{k}\ram(a,b;\omega,\omega^m z)}{(a;\omega)_{k}(\omega^maz/b;\omega)_{k}\ram(a,b;\omega,z)}\Rightarrow\\
\label{eq:awk-zwm-terated}\frac{\ram(\omega^ka,b;\omega, \omega^m z)}{\ram(a,b;\omega,z)}&=\frac{(a/b;\omega)_{k}(z;\omega)_{m}}{b^m(a;\omega)_{k}(\omega^maz/b;\omega)_{k}(az/b;\omega)_{m}}.
\end{align}
Cyclic reindexing gives
\begin{equation}\label{eq:psi-shift-a-b}
 \ram(\omega^l a,\omega^l b;\omega, z)
 =\frac{(\omega b;\omega)_l}{z^l(a;\omega)_l}
  \ram(a,b;\omega, z).
\end{equation}
In~\eqref{eq:psi-shift-a-b}, replacing $a$ by $\omega^{k-l}a$ and $z$ by $\omega^m z$, and then using~\eqref{eq:awk-zwm-terated}, we obtain
\begin{align}
 \nonumber \ram(\omega^k a,\omega^l b;\omega, \omega^m z)
 =\frac{(\omega b;\omega)_l}{\omega^{lm}z^l(\omega^{k-l}a;\omega)_l}
  \ram(\omega^{k-l}a,b;\omega, \omega^{m}z)\Rightarrow\\
 \nonumber \frac{\ram(\omega^k a,\omega^l b;\omega, \omega^m z)}{\ram(a,b;\omega,z)}=\frac{(\omega b;\omega)_l}{\omega^{lm}z^l(\omega^{k-l}a;\omega)_l}
 \frac{(a/b;\omega)_{k-l}(z;\omega)_{m}}{b^m(a;\omega)_{k-l}(\omega^maz/b;\omega)_{k-l}(az/b;\omega)_{m}}\Leftrightarrow\\
\label{eq:awk-bwl-zwm-iterated} \frac{\ram(\omega^k a,\omega^l b;\omega, \omega^m z)}{\ram(a,b;\omega,z)}=
 \frac{(\omega b;\omega)_l(a/b;\omega)_{k-l}(z;\omega)_{m}}{\omega^{lm}z^lb^m(a;\omega)_{k}(az/b;\omega)_{k-l+m}}.
\end{align}
Formula~\eqref{eq:awk-bwl-zwm-iterated} is precisely~\eqref{eq:H-master} for positive $k-l,l,m$. The negative shifts follow
from the reciprocal Pochhammer convention and the $N$-periodicity.
\end{proof}
\end{document}